\documentclass[a4paper,11pt]{article}
\pdfoutput=1 

\usepackage{jcappub} 

\usepackage[T1]{fontenc} 
\usepackage{graphicx}
\usepackage{epsfig}
\usepackage{rotate}
\usepackage{amsmath}
\usepackage{amssymb}
\usepackage{amsfonts}
\usepackage{bm}
\usepackage{enumerate}
\usepackage{afterpage}
\usepackage{xcolor}
\usepackage{natbib, hyperref}
\usepackage[normalem]{ulem}
\usepackage[makeroom]{cancel}

\UseRawInputEncoding

\newcommand{\ud}{\mathrm{d}}
\newcommand{\p}{\partial}
\newcommand{\cH}{\mathcal{H}}
\newcommand{\HH}{\mathcal{H}}
\newcommand{\Q}{\mathcal{Q}}

\def\be{\begin{equation}}
\def\ee{\end{equation}}
\def\bea{\begin{eqnarray}}
\def\eea{\end{eqnarray}}
\def\fnl{f_{\rm NL}}

\def\n{\bm{n}}
\def\k{\bm{k}}
\def\v{\bm{v}}
\def\x{\bm{x}}

\def\degree{^\circ}

\def\in{{\rm in}}
\def\on{(1)}
\def\tw{(2)}

\newcommand{\red}[1]{{\color{red}{#1}}}
\newcommand{\blue}[1]{{\color{magenta}{#1}}}
\newcommand{\bro}[1]{{\color{black}{#1}}}

\title{{Local primordial non-Gaussianity in the relativistic galaxy bispectrum}}

\author{Roy Maartens$^{1,2}$, Sheean Jolicoeur$^1$, Obinna Umeh$^2$,\\  Eline M. De Weerd$^3$, Chris Clarkson$^{3,1}$}
\affiliation{$^1$Department of Physics \& Astronomy, University of the Western Cape, Cape Town 7535, South Africa\\
$^{2}$Institute of Cosmology \& Gravitation, University of Portsmouth, Portsmouth PO1 3FX, UK\\
$^{3}$School of Physics \& Astronomy, Queen Mary University of London, London E1 4NS, UK}

\abstract{
Next-generation galaxy and 21cm intensity mapping surveys will rely on a combination of the power spectrum and bispectrum for high-precision measurements of primordial non-Gaussianity. In turn, these measurements will allow us to distinguish between various models of inflation. However, precision observations require theoretical precision at least at the same level. We extend the theoretical understanding of the galaxy bispectrum by incorporating a consistent general relativistic model of galaxy bias at second order, in the presence of local primordial non-Gaussianity. {The influence of primordial non-Gaussianity on the bispectrum extends beyond the  galaxy bias and the dark matter density, due to redshift-space effects. The standard redshift-space distortions at first and second order produce a well-known primordial non-Gaussian  imprint on the bispectrum. Relativistic corrections to redshift-space distortions
generate new contributions to this primordial non-Gaussian signal, arising from: (1)~a coupling of first-order scale-dependent bias with first-order relativistic observational effects, and (2)~linearly evolved non-Gaussianity in the second-order velocity and metric potentials which appear in relativistic observational effects.}
Our analysis allows for a consistent separation of the
 relativistic  `contamination' from the primordial signal, in order to avoid biasing the measurements by using an incorrect theoretical model. We show that the bias from using a Newtonian analysis of the squeezed bispectrum could be $\Delta \fnl\sim 5$ for a Stage IV H$\alpha$ survey.}

\begin{document}
\maketitle
\date{\today}
\flushbottom
\newpage
\section{Introduction}
Galaxy number counts are distorted by projection effects that arise from observing on the past lightcone. {The dominant perturbative effect on sub-Hubble  scales is from redshift-space distortions (RSD) \cite{Sargent:1977,Kaiser:1987qv}, 
which constitute the standard Newtonian approximation to projection effects. Lensing magnification produces the best-known relativistic correction to RSD \cite{Villumsen:1995ar}, but there are further relativistic effects \cite{Yoo:2009au,Yoo:2010ni,Challinor:2011bk,Bonvin:2011bg}.
The basic idea is the following. The number of sources, $\ud \mathbb{N} $,  {above the luminosity threshold} that are counted by the observer in a solid angle element about unit direction $\n$ and in a redshift interval about a central redshift $z$, is given by 
\be \label{dn}
\ud \mathbb{N} =N_g\, \ud z\,\ud\Omega_{\n}=  n_g\, \ud {\cal V} \,. 
\ee
The second equality relates the observed quantities to those measured in the rest frame of the source.
$N_g$ is the number that is counted by the observer per redshift per solid angle, while 
$ n_g$ is the number per proper volume, which is not observed by the observer but is the quantity that would be measured at the source. Similarly,  $\ud {\cal V}$ is not the observed volume element but the corresponding proper volume element at the source.  

Then the observed number density contrast, $\Delta_g=(N_g-\bar{N}_g)/\bar{N}_g$, is related to the proper number density contrast at the source, $\delta_g=(n_g-\bar{n}_g)/\bar{n}_g$, by volume, redshift and {luminosity} perturbations. 
At first order  in Poisson gauge,  the gauge-independent relation \eqref{dn} leads to
\bea
 \Delta_g&=&\delta_g +\,\mbox{RSD +  lensing effect + other relativistic effects} \notag\\
\label{obdel}
&=& \delta_g-{1\over {\cal H}}\n \cdot \bm\nabla \big(\v  \cdot \n \big) + 2(1-{\cal Q})\kappa +  A\big(\v  \cdot \n \big)+B \Psi + \int\! \ud \chi\,C\Psi' +\int\! \ud \chi\,E\Psi \,.
\eea
Here ${\cal H}=\ud\ln a/\ud \eta=(\ln a)'$ is the conformal Hubble rate, $\v=\bm{\nabla} V$ is the peculiar velocity ($V$ is not to be confused with the often-used alternative $v=|\v|$), $\kappa$ is the integrated lensing convergence,  ${\cal Q}$ is the magnification bias, $\chi$ is the comoving line-of-sight distance and the integrals are from source to observer. 
The perturbed metric is given by
\be \label{pds}
a^{-2}\ud s^2=-\big(1+2\Phi \big)\ud \eta^2+ \big(1-2\Psi \big)\ud \x^2\,,
\ee
and we have assumed $\Phi=\Psi$. The time-dependent factors $A,B,C,E$ in \eqref{obdel} correspond respectively to Doppler, Sachs-Wolfe, integrated Sachs-Wolfe and time-delay effects. In Fourier space the Doppler term scales as $\partial V \propto ({\cal H}/k)\delta_m$, while the remaining terms scale as $\Psi \propto ({\cal H}/k)^2\delta_m$. Thus the other relativistic effects are suppressed on sub-Hubble scales, unlike the lensing effect, which scales as $\partial^2 \Psi \propto \delta_m$.

{The case of 21cm intensity mapping follows from the number count expressions  by using the `dictionary' given in \cite{Hall:2012wd,Alonso:2015uua,Fonseca:2015laa} at first order and in \cite{Umeh:2015gza,DiDio:2015bua,Jolicoeur:2020eup} at second order.}

The physical definition of linear Gaussian galaxy bias is in the joint matter-galaxy rest frame, which corresponds to the comoving gauge (`C gauge'),\footnote{In the $\Lambda$CDM model  the comoving and synchronous gauges coincide.} so that (omitting luminosity dependence for brevity), 
\be \label{b1}
\delta_{g{\rm C}}(a,\x)=b_1(a)\delta_{m{\rm C}}(a,\x)\,.
\ee 
This relation is gauge-independent because C gauge corresponds to the physical rest frame.
When transforming to other gauges, $\delta_g$ is in general no longer proportional to $\delta_m$ \cite{Challinor:2011bk,Bruni:2011ta,Jeong:2011as}. For example,  in the Poisson gauge of  \eqref{obdel} and~\eqref{pds}, 
\be \label{b1pg}
\delta_g =b_1\delta_{m{\rm C}} + (3-b_e){\cal H}V\,,\quad b_e= {\partial \ln (a^3\bar{n}_g) \over \partial \ln a}\,,
\ee
where $b_e$ is known as the evolution bias, which encodes the non-conservation of the background comoving galaxy number density. The velocity potential $V$ scales as $\Psi$ by the {Euler} equation, $ V\propto  \Psi \propto ({\cal H}/k)^2\delta_m$, and therefore
the gauge correction $(3-b_e){\cal H}V$ is only non-negligible on Hubble scales and may be neglected in a Newtonian approximation. 

Local  primordial non-Gaussianity (PNG) generates scale-dependent linear bias, with constant parameter $\fnl$ 
\cite{Dalal:2007cu,Matarrese:2008nc}:
\be \label{b1ng}
 b_{1}(a)~\to~ b_1(a)+3\,\delta_{\rm crit}\Omega_{m0}H_0^2 \,{\big[b_1(a)-1 \big]\over D(a)}\, 
 g_{\in}\,{\fnl \over T(k)k^2}\,.
\ee
The threshold density contrast for collapse is usually taken to be $\delta_{\rm crit}=1.686$, and the growth factor $D$  is normalised to 1 today ($a_0=1$), i.e. $\delta_m(a,\k)=D(a)\delta_{m0}(\k) $. The growth suppression factor for the potential $\Psi$ is $g=D/a$, which is thus also normalised as $g_0=1$, with initial value $g_{\in}$ deep in the matter era,  and $T$ is the  transfer function. Note that \eqref{b1ng} follows the CMB convention
for $\fnl$ \cite{Baldauf:2010vn,Desjacques:2016bnm};  $g_{\in}$ can be removed from \eqref{b1ng} if $D$ is normalised as $D_{\in}=a_{\in}$.  
In a $\Lambda$CDM model we have the useful relation \cite{Villa:2015ppa}
\be \label{ggin}
{g_{\in} \over g}={3\over5}\Big(1+ {2f \over 3\Omega_m} \Big),
\ee
where the growth rate of linear matter perturbations, $f=\ud \ln D/\ud \ln a$, is very well approximated by $f(a)=\Omega_m(a)^{0.545}$. 

The PNG component of galaxy bias in \eqref{b1ng} 
scales as $H_0^2/k^2$ on ultra-large scales, i.e. above the equality scale, $k< k_{\rm eq}$, where $T\approx 1$. It
is strongly suppressed on scales $k\gg k_{\rm eq}$ by $T(k)$. PNG has a similar impact on the power spectrum to the impact of ultra-large-scale relativistic effects. This means that relativistic effects contaminate the primordial signal --  leading to  biases if a Newtonian approximation is used to model the galaxy power spectrum (see \cite{Bruni:2011ta, Jeong:2011as, Camera:2014sba}). 
The relativistic galaxy power spectrum has been used to analyse and predict the capability of future galaxy and intensity mapping surveys to
measure the local PNG parameter $\fnl$, while avoiding the bias  that is inherent in a Newtonian analysis (see e.g. \cite{Bruni:2011ta, Jeong:2011as, LopezHonorez:2011cy, Yoo:2012se, Raccanelli:2013dza,Camera:2014bwa, Camera:2014sba, Raccanelli:2015vla, Alonso:2015uua, Alonso:2015sfa, Fonseca:2015laa, Fonseca:2016xvi, Abramo:2017xnp, Lorenz:2017iez, Fonseca:2018hsu, Ballardini:2019wxj, Grimm:2020ays, Bernal:2020pwq, Wang:2020ibf}).  

The tree-level bispectrum requires the number counts
in redshift space up to second order.
In the Newtonian approximation, the projection effects are the second-order RSD terms (see e.g. \cite{Tellarini:2016sgp}). The relativistic corrections
 to RSD at second-order  are extremely complicated, since they involve quadratic couplings of all the first-order terms, as well as introducing new terms that do not enter at first order, such as the {transverse peculiar velocity}, the lensing deflection angle and the lensing shear  \cite{Bertacca:2014dra, Bertacca:2014wga, Yoo:2014sfa, DiDio:2014lka, Bertacca:2014hwa}.   {There are further relativistic corrections that are not projection effects. Firstly,
 the Newtonian model of second-order galaxy bias in the comoving frame requires a relativistic correction, unlike the first-order bias (see Section \ref{sdb}). Secondly,
 and similar to the first-order case, the second-order galaxy bias relation needs  relativistic gauge corrections  when using non-comoving gauges such as the Poisson gauge. These are second-order extensions of equations like \eqref{b1pg}.}  
{In summary, the second-order relativistic corrections to the galaxy bispectrum in the Gaussian case are: 
\begin{itemize}
\item
relativistic projection corrections to the Newtonian RSD  \cite{Bertacca:2014dra, Bertacca:2014wga, Yoo:2014sfa, DiDio:2014lka, Bertacca:2014hwa};
\item
relativistic corrections to the Newtonian bias model in the comoving frame at second order, which were only recently derived  \cite{Umeh:2019qyd, Umeh:2019jqg};
\item  
 relativistic gauge corrections to the second-order number density when using non-comoving gauges   \cite{Bertacca:2014dra, Bertacca:2014wga}. 
 \end{itemize}}

As in the case of the power spectrum, local PNG affects the bispectrum on very large scales, which is also where the relativistic effects are strongest. This leads again to a contamination of the primordial signal by relativistic effects, necessitating a relativistic analysis. A Gaussian primordial universe could be mistakenly interpreted as non-Gaussian if a Newtonian model is used for the bispectrum in analysis of the data, as shown by 
\cite{Kehagias:2015tda,Umeh:2016nuh,Jolicoeur:2017nyt,Koyama:2018ttg}. 

{There are important differences between the power spectrum and bispectrum:
\begin{itemize}
\item
At first order, there is no relativistic correction to the bias model in comoving gauge -- the relativistic correction arises at second order \cite{Umeh:2019qyd, Umeh:2019jqg}. Therefore the tree-level bispectrum contains a relativistic correction to the bias model, but the tree-level power spectrum does not.
\item
There is no PNG signal in the primordial {\em matter} power spectrum at tree level, so that the local PNG signal in the tree-level galaxy power spectrum is sourced only by scale-dependent bias. 
\item
By contrast, local PNG in the galaxy bispectrum is sourced by scale-dependent bias, by the primordial matter bispectrum {and by RSD at second order (see \cite{Tellarini:2016sgp} and Section \ref{ss-mvp} below)}.
\item
 Second-order relativistic {corrections to RSD}  induce new local PNG effects in the bispectrum, via~(1) a coupling of first-order scale-dependent bias to first-order relativistic projection effects,  and {(2)~the linearly evolved PNG in second-order velocity and metric potentials, which appear in relativistic projection effects (absent in the standard Newtonian analysis)}. 
\end{itemize}}

Since local PNG affects the power spectrum and bispectrum differently, a Newtonian analysis could mistakenly identify inconsistencies between the power spectrum and bispectrum $\fnl$ measurements, which could wrongly lead to an inference of hidden systematics or deviations from general relativity.}

PNG in the galaxy bispectrum has been extensively investigated in the Newtonian approximation. Most work has used the Fourier bispectrum, implicitly incorporating a plane-parallel assumption (see e.g. 
\cite{Verde:1999ij,Scoccimarro:2003wn,Sefusatti:2006pa,Sefusatti:2007ih,Giannantonio:2009,Baldauf:2010vn,Tellarini:2015faa,Tellarini:2016sgp,Desjacques:2016bnm,Watkinson:2017zbs,Majumdar:2017tdm,Karagiannis:2018jdt,Yankelevich:2018uaz,Sarkar:2019ojl,Karagiannis:2019jjx,Bharadwaj:2020wkc,Karagiannis:2020dpq, MoradinezhadDizgah:2020whw}) and we follow this approximation.
Our previous work \cite{Umeh:2016nuh} included the local (non-integrated) relativistic effects in the Fourier bispectrum for the first time. This was extended by our work \cite{Jolicoeur:2017nyt, Jolicoeur:2017eyi, Jolicoeur:2018blf,Clarkson:2018dwn,Maartens:2019yhx,deWeerd:2019cae, Jolicoeur:2020eup, Umeh:2020xxx}, all in the case of primordial Gaussianity. 
Here we incorporate local PNG into the relativistic bispectrum. This involves applying the recent results of \cite{Umeh:2019qyd, Umeh:2019jqg} on relativistic corrections to the second-order galaxy bias model. {In addition, we derive the new local  PNG terms induced by  a coupling of first-order scale-dependent bias and first-order relativistic projection effects and by linearly evolved second-order relativistic projection effects.}

The paper is structured as follows. Section \ref{ss-mvp} reviews the relativistic correction to the 
galaxy bias, including  the case of local PNG. In addition, we show how the linearly evolved second-order metric and velocity potentials carry a primordial non-Gaussian signal, which is imprinted in the bispectrum by relativistic projection effects. In Section \ref{sec3}, after presenting the relativistic correction to the matter bispectrum, we discuss the number density contrast in redshift space, which brings into play the relativistic projection effects. We combine the various results to derive the relativistic galaxy bispectrum, including all local PNG effects, and we show examples of the galaxy bispectrum for a Stage IV H$\alpha$ spectroscopic survey. We summarise and conclude in Section \ref{sec4}.

~\\ \noindent {\bf\em Conventions used:} 

We assume  a flat $\Lambda$CDM model, based on general relativity and perturbed up to second order, in which the matter is pressure-free and irrotational on perturbative scales. Generalisations to allow dynamical dark energy and relativistic modified gravity are straightforward, but are not included. For numerical calculations, we use the Planck 2018 best-fit parameters \cite{Aghanim:2018eyx}.
Perturbed quantities are expanded as
$X+X^{\tw}/2$, and may be split as  $X_{\rm N}+X_{\rm GR}+X_{\rm nG}$, 
and similarly at second order, where N denotes the Newtonian approximation, GR denotes the relativistic correction and  nG denotes the local PNG contribution. {GR corrections are highlighted in \blue{magenta}.}

Our definition of the metric potentials in \eqref{pds} leads to the first-order Poisson equation 
\be \label{pe1}
\nabla^2 \Psi=+{3\over2}\Omega_m{\cal H}^2\,\delta_{\rm C} \,,
\ee 
where  $ \Phi= \Psi$ in $\Lambda$CDM. Here and in the remainder of the paper, we omit the subscript $m$ on the matter density contrast for brevity.
At second order, the perturbed metric in Poisson gauge is given by
\be \label{pds2}
a^{-2}\ud s^2=-\big[1+2\Psi + \Phi^{\tw}\big]\ud \eta^2+ \big[1-2\Psi -  \Psi^{\tw}\big]\ud \x^2\,.
\ee
\bro{Here we have neglected the relativistic vector and tensor modes that are generated by scalar mode coupling, so that we only consider the relativistic scalar contribution to the bispectrum. This approximation is justified by the fact that the relativistic vector contribution to the bispectrum is typically 2 orders of magnitude below the relativistic scalar contribution on observable scales, while the relativistic tensor contribution is typically an order of magnitude below that of the vector contribution (see \cite{Jolicoeur:2018blf}).}
\\


\section{Local primordial non-Gaussianity in the galaxy bias}
\label{sdb}

Local PNG is defined as a simple form of nonlinearity in the primordial curvature perturbation, which is local in configuration space. In terms of the gravitational potential {deep in the matter era}, we have
\begin{equation}
-\Big[\Psi_{\in}(\x) +{1\over2} \Psi^{\tw}_{\in}(\x)\Big] = \varphi_{\in}(\x) + \fnl\big[\varphi_{\in}(\x)^{2}  - \big\langle \varphi_{\in}^{2} \big\rangle\big],
\label{e1.1} 
\end{equation}
where $\varphi_{\in}$ is the first-order Gaussian part.
The standard definition of $\fnl$ uses a convention for $\Psi$ that is different to ours, with a minus on the right of the Poisson equation \eqref{pe1}. In order to keep the standard sign of $\fnl$, we made a sign change on the left of \eqref{e1.1}.
($\fnl$  in  \cite{Villa:2015ppa,Koyama:2018ttg, Umeh:2019jqg} is  of opposite sign to the standard sign that we use.) 

\subsection{First-order bias}

In \eqref{e1.1}, the Gaussian part of the potential deep in the matter era {(but after decoupling)} is {related to the linear primordial potential by the transfer function:
\be \label{vpe}
\varphi_{\in}(\k)=T(k)\,\varphi_{\rm p}(\k) {\quad \mbox{for}\quad a_{\rm p}\ll a_{\rm eq} \ll a_{\in} \,. }
\ee
{Here $\varphi_{\rm p}(\k)=-9
\Psi(a_{\rm p},\k)/10$, where  the factor 9/10 ensures conservation of the curvature perturbation on super-Hubble scales.}
After equality, the potential evolves with the growth suppression factor, so that
\be\label{pottp}
{\varphi(a,\k)={g(a) \over g_{\in}}  \varphi_{\in}(\k) \quad \mbox{for}\quad a\geq a_{\in}> {a_{\rm dec}}\,.}
\ee
\begin{figure}[!h]
\centering
\includegraphics[width=.49\textwidth, angle=0]{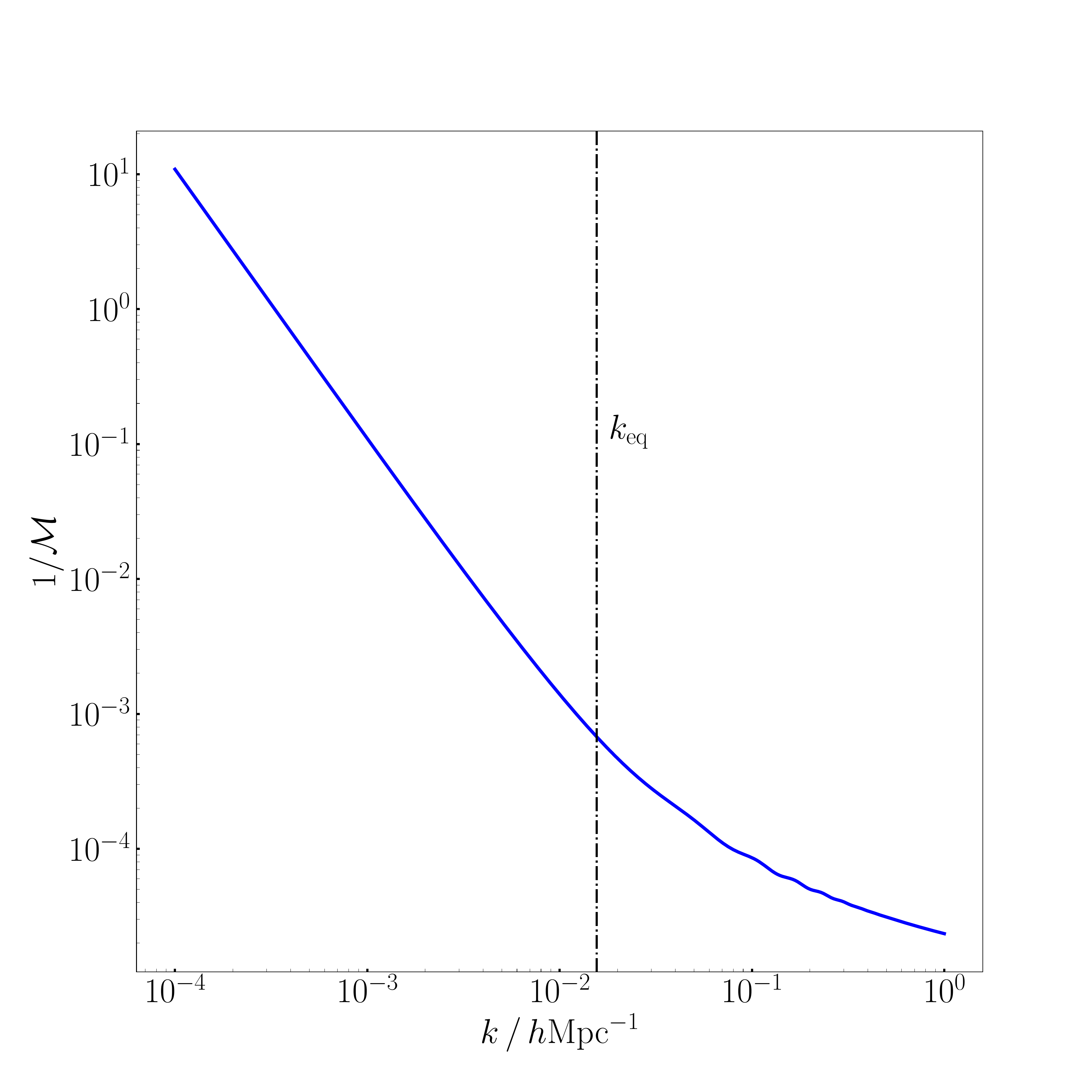}
\caption{{${\cal M}^{-1}= \varphi_{\rm p}/\delta^{\on}_{\rm C}$ at $z=1$.}}
\label{mplot}
\end{figure}

We relate the late-time matter density contrast  to the primordial potential via the Poisson equation \eqref{pe1}, using \eqref{ggin}, \eqref{vpe} and \eqref{pottp}:
\bea \label{alpha}
\delta_{\rm C}(a,\k)=  {\cal M} (a,k)  \varphi_{\rm p} (\k) \quad \mbox{where} \quad 
{{\cal M} (a,k)  = {10 \over 3\cH(a)^2\big[3\Omega_m(a) +2f(a) \big]}\,k^2 \, T(k)}\,.
\eea
This relation is illustrated in Fig. \ref{mplot}. 
{The matter and number density contrasts can be written as 
\be \label{dcng}
\delta_{\rm C}= \delta_{\rm C,N} \quad \mbox{and}\quad
\delta_{g\rm C}= \delta_{g\rm C,N}+ \delta_{g\rm C,nG} \,.
\ee
This follows since there is no GR correction to either contrast and no PNG in the Gaussian matter density contrast:}
\be \label{dgro1}
\blue{\delta_{{\rm C,GR}} =0=\delta_{g{\rm C,GR}}}\,,\quad \delta_{{\rm C,nG}} =0\,.
\ee
Then it follows that}
\bea\label{dg1b}
\delta_{g{\rm C}}= \delta_{g \rm C,N}+ \delta_{g{\rm C,nG}}=   b_{10}\,\delta_{\rm C}+ b_{01}\,\varphi_{\rm p}\,, 
\eea
where the Gaussian and non-Gaussian  bias coefficients are 
\bea 
\label{bias1}
b_{10}= b_1\,,\quad
b_{01}= 2\fnl \delta_{\rm crit}(b_{10}-1)\,.
\eea
The relations \eqref{alpha}--\eqref{bias1} then recover \eqref{b1ng}.

{At first order,  there is {\em no} GR correction to the  bias relation  expressed in the matter-galaxy rest frame. This is no longer true at second order.

The first-order metric  potential is Gaussian by \eqref{e1.1} and has no GR correction by \eqref{dgro1} and  the Poisson equation. From the Euler equation ($V'+{\cal H}V=- \Psi$) it follows that the velocity also has no GR and no PNG corrections:
\be \label{pvo1}
\Psi=\Psi_{\rm N}\,,\qquad V=V_{\rm N}\,.
\ee 


\subsection{Second-order bias: {Newtonian} approximation}

At second order, the  galaxy bias is {physically} defined in comoving gauge, {but any gauge may be used in general relativity. Standard Newtonian perturbation theory is often given in an Eulerian frame, and so it is useful for comparison}  to express the  bias in a suitable Eulerian frame. We use Poisson gauge here, following \cite{Umeh:2016nuh,Tram:2016cpy, Jolicoeur:2017nyt, Jolicoeur:2017eyi, Jolicoeur:2018blf,Clarkson:2018dwn,Maartens:2019yhx,deWeerd:2019cae}, but with the galaxy and matter density contrasts in total-matter gauge (`T gauge').}  {The total-matter gauge is a convenient  Eulerian choice  for the density contrasts, since it has
the same spatial coordinates as the Poisson gauge at first order and the same time-slicing as the comoving gauge at first and second orders  \cite{Bartolo:2015qva,Villa:2015ppa,Tram:2016cpy}. 
As a result, at first order
the total-matter density contrasts coincide with those of the comoving gauge: 
$\delta_{{\rm T}}=\delta_{{\rm C}}$, $\delta_{g{\rm T}}=\delta_{g{\rm C}}$, 
and we can rewrite  \eqref{dg1b} as 
\bea
\delta_{g{\rm T}} ={\delta_{g \rm T,N}+\delta_{g{\rm T,nG}}} 
=b_{10}\,\delta_{\rm T}+ b_{01}\,\varphi_{\rm p}={\Big(b_{10} + {b_{01} \over {\cal M}} \Big) \delta_{\rm T}}\,.
 \label{dg1bt}
\eea

At second order, the total-matter and Poisson matter density contrasts agree in the Newtonian approximation: 
$\delta_{\rm T,N}^{\tw}=\delta^{\tw}_{\rm N}$, while
the comoving and total-matter Newtonian density contrasts are related via a purely spatial gauge transformation \cite{Bertacca:2015mca,Villa:2015ppa, Jolicoeur:2017nyt,Umeh:2019qyd}: 
\be \label{ttc2}
{\delta_{\rm T,N}^{\tw}} = {\delta_{\rm C,N}^{\tw}}+2\xi^i\partial_i \delta_{{\rm C}} \,,\quad {\delta_{g\rm T,N}^{\tw}} = {\delta_{g\rm C,N}^{\tw}}+2\xi^i\partial_i \delta_{{g\rm C}} \,, 
\ee
where
\be \label{xi}
\xi^i = \partial^i \nabla^{-2}\delta_{{\rm C}} = \partial^i \nabla^{-2}\delta_{{\rm T}} \,.
\ee
(The GR parts of the second-order density contrasts in comoving and total-matter gauges are equal; see below.)

{For the small scales involved in local clustering of matter density, the Poisson equation at second order has the same Newtonian form as at first order.
 {Then we can extend \eqref{alpha} up to second order to define the linearly evolved local PNG part of the density contrast, whose nonlinearity is purely primordial: 
\be \label{alpha1a}
\delta_{\rm T,nG}^{\tw}={\cal M} \,\varphi_{\rm p}^{(2)} = 2\fnl  \,{\cal M} \,\varphi_{\rm p}*\varphi_{\rm p}\,,
\ee 
where the $*$ denotes a convolution in Fourier space.
This leads to
\be\label{alpha2}
\delta_{\rm T,nG}^{\tw}= 2\fnl \,{\cal M} (a,k)\int {\ud \k' \over (2\pi)^3}\, {\delta_{\rm T}(a,\k')\over {\cal M} (a,k')}\,{\delta_{\rm T}(a,\k-\k') \over {\cal M} (a,|\k-\k'|)}\,.
\ee

In order to include the nonlinearity due to gravitational evolution, we add the standard Newtonian  contribution for Gaussian initial conditions to the local PNG part:}  
 \bea \label{alpha3}
&& \delta_{\rm T,{N}}^{\tw}(a,\k) + \delta_{\rm T,nG}^{\tw}(a,\k)  
\\ \notag 
&&{}= \int {\ud \k' \over (2\pi)^3}\, \bigg[{F_{2}}(a,\k',\k-\k') +  2\fnl \,{{\cal M} (a,k) \over
{\cal M} (a,k')\, {\cal M} (a,|\k-\k'|)} \bigg] \delta_{\rm T}(a,\k')\,\delta_{\rm T}(a,\k-\k') \,.
\eea 
The standard Newtonian mode-coupling kernel
 for $\Lambda$CDM is \cite{Villa:2015ppa}:
\bea
F_{2}(a,\bm{k}_{1},\bm{k}_{2}) &=&1+{F(a)\over D(a)^2}+ \Big({k_1\over k_2}+ {k_2\over k_1}\Big)\hat{\k}_1\cdot \hat{\k}_2 +\bigg[1-{F(a)\over D(a)^2}\bigg]\big( \hat{\k}_1\cdot \hat{\k}_2\big)^2\,,\label{f2k}
\eea
where $F$ is the second-order growth factor.  The Einstein--de Sitter relation $F/D^2=3/7$ is a very good approximation in $\Lambda$CDM. We use this approximation, in which $F_2$ is effectively time independent.}

At second order, the standard Newtonian bias model,  including tidal bias in the Gaussian part and all local PNG contributions, is given by  (see   \cite{Desjacques:2016bnm} for a comprehensive treatment):
\bea
\delta^{\tw}_{g{\rm T,{N}}} +\delta_{g\rm T,nG}^{\tw}&=& b_{10}\, {\delta^{\tw}_{{\rm T,{N}}}}+ b_{20} \big(\delta_{{\rm T}}\big)^2   + b_s\,s^2 
\notag \\
&&{} +b_{10}\, \delta_{\rm T,nG}^{\tw} + b_{11}\, \delta_{{\rm T}}\,\varphi_{\rm p} +  b_{{n}}\, \xi^i\,\partial_i\,\varphi_{\rm p}+b_{02} \big(\varphi_{\rm p}\big)^2\,. \label{dg2bn}
\eea
The (Eulerian) bias parameters in the case of Gaussian initial conditions are in the first line on the right-hand side:  the  linear and quadratic biases, $b_{10}$ and $b_{20}$, and the tidal bias $b_s$, where  
\be \label{ts}
s^2=s_{ij}s^{ij}\,, \quad s_{ij} =\Big(\partial_i \partial_j-{1\over3}\delta_{ij}\nabla^2 \Big){\nabla^{-2}\delta_{\rm T}} 
\,.
\ee
The second line of \eqref{dg2bn} contains the local PNG contribution, with three new bias parameters $b_{11}, b_{{n}}, b_{02}$. The first term is the {primordial dark matter contribution}, from \eqref{alpha3}; note that $\tilde\delta^{\tw}_{{\rm T,N}}$ is proportional to $\fnl$.
{The $b_{11}, b_n$ terms scale as $(\cH^2/k^2)\,(\delta_{\rm T})^2$, while the $b_{02}$ term is $\mathcal{O}(\cH^4/k^4)$.}
 The new bias parameters vanish when $\fnl=0$; in the presence of local PNG, they are given by \cite{Tellarini:2015faa, Desjacques:2016bnm,Umeh:2019jqg}: 
\bea
b_{11} &=& 4\fnl \Big[\delta_{\rm crit}\,b_{20}+\Big({13\over21}\delta_{\rm crit}-1 \Big) (b_{10}-1)+1\Big] 
\,, \label{b11}
\\
b_{{n}} &=& 4\fnl \Big[ \delta_{\rm crit}(1-b_{10}) +1\Big] \,,  \label{bxi}
\\
b_{02} &=& 4\fnl^2 \delta_{\rm crit}\Big[\delta_{\rm crit}\,b_{20}-2\Big({4\over21}\delta_{\rm crit}+1 \Big) (b_{10}-1)\Big]. \label{b02}
\eea
Note that the expressions for the bias coefficients in \eqref{b11}--\eqref{b02}, as well as for $b_{01}$ in \eqref{bias1}, are based on a universal halo mass function. {(For recent work on the limits of the universality assumption, see  \cite{Barreira:2020kvh,Barreira:2020ekm}.)} \\

\subsection{Second-order bias: {relativistic corrections}}
~\\
     
{The relativistic second-order galaxy bias model has been derived in
\cite{Umeh:2019qyd} (Gaussian case) and \cite{Umeh:2019jqg} (with local PNG). {The key feature to bear in mind is the following:
 \begin{quote} 
{\em GR corrections in the galaxy number density contrast $\delta^{\tw}_{g{\rm T}}$ do not change the galaxy bias terms in \eqref{dg2bn}, which contain all the local PNG effects.}
\end{quote}
This separation between GR effects and local PNG in the number density can be understood as follows. 
\begin{itemize}
\item
The intrinsic  nonlinearity of GR  modulates the galaxy number density  via large-scale modes. 
However,  this does not affect small-scale clustering: GR effects do {\em not} modulate the variance of small-scale density modes {\cite{Koyama:2018ttg,Dai:2015jaa,dePutter:2015vga}}.
\item
By contrast, local PNG imprints a primordial long-short coupling that induces a long-mode modulation of the variance and thus changes the galaxy bias.
\end{itemize}
As a consequence, we expect that relativistic corrections to the bias relation should be independent of  non-Gaussianity and apply only on ultra-large scales  (for a different view, see  \cite{Matarrese:2020why}). 
These two features are consistent with the behaviour of  \eqref{dg2bn} under change of gauge:
 \begin{quote} 
{\em The Newtonian bias relation \eqref{dg2bn} is gauge-independent only on small scales.\\
Relativistic
corrections to \eqref{dg2bn}  are needed to enforce gauge-independence of the bias relation on ultra-large scales.}
\end{quote}

As shown in  \cite{Umeh:2019qyd,Umeh:2019jqg}, gauge-independence requires the addition to  \eqref{dg2bn} of
 the relativistic part of the second-order matter density contrast. The relativistic modes are super-Hubble at equality and arise from nonlinear GR  corrections to the Newtonian Poisson equation   \cite{Bruni:2013qta,Bartolo:2015qva,Villa:2015ppa,Tram:2016cpy}: 
\bea\label{dtgr2}
\blue{\delta_{\rm C,GR}^{\tw}=\delta_{\rm T,GR}^{\tw}=
{20\over 3}\, \delta_{{\rm T}}\,{\hat{\varphi}_{\rm in}} 
-{5\over3}\, \xi^i\,\partial_i \, {\hat\varphi_{\rm in}} \equiv \delta^{\tw}_{g{\rm T,GR}}}\,. 
\eea
Here $\hat\varphi_{\rm in}$ is the ultra-large scale potential deep in the matter era, 
\bea \label{hatv}
 \hat\varphi_{\rm in}(\k) =  \varphi_{\rm in}\big(\k\, | \, k<k_{\rm eq}\big) \,.
\eea
When we relate $\hat\varphi_{\rm in}$ to the density contrast today, via \eqref{vpe} and \eqref{alpha}, we need to impose $T=1$ on the transfer function, by \eqref{hatv}.}

The relativistic second-order galaxy bias model of
\cite{Umeh:2019jqg} can  be written in T-gauge as
\bea
\delta^{\tw}_{g{\rm T}} &=& \delta^{\tw}_{g{\rm T,N}}+ \delta_{g\rm T,nG}^{\tw} +\blue{\delta^{\tw}_{g{\rm T,GR}}}
\,,
\label{dg2b1}
\eea
where
\bea
\delta^{\tw}_{g{\rm T,N}} &=& b_{10}\, \delta^{\tw}_{{\rm T,N}}+ b_{20} \big(\delta_{{\rm T}}\big)^2   + b_s\,s^2 \,,
\label{dg2b2}\\
\delta_{g\rm T,nG}^{\tw} &=& b_{10}\,  \delta_{\rm T,nG}^{\tw}
 + b_{11}\, \delta_{{\rm T}}\,\varphi_{\rm p} +  b_{{n}}\, \xi^i\,\partial_i\,\varphi_{\rm p}+b_{02} \big(\varphi_{\rm p}\big)^2,
\label{dg2b3} \\
\blue{\delta^{\tw}_{g{\rm T,GR}}}&\blue{=}& \blue{{20\over 3}\, \delta_{{\rm T}}\,\hat\varphi_{{\rm in}} - {5\over3}\, \xi^i\,\partial_i\,\hat\varphi_{{\rm in}}} \,. \label{dg2b}
\eea
Here \eqref{dg2b2} and \eqref{dg2b3} recover the Newtonian relation  \eqref{dg2bn}.  

{Both the local PNG and GR terms scale as  $(\cH^2/k^2)\,(\delta_{\rm T})^2$, {so that the GR correction {\em cannot} be neglected.
Although they are of the same order of magnitude, there is a key distinction between them:} 
 {local PNG induces a short-long mode coupling, and thus affects the primordial potential $\varphi_{\rm p}$ on small scales, while the GR corrections affect only the ultra-large-scale primordial  modes.} In the absence of local PNG, i.e. for $\fnl=0$, the GR terms  survive
and constitute the relativistic bias correction in the case of Gaussian initial conditions, as derived in \cite{Umeh:2019qyd}.

Finally, we transform \eqref{dg1bt} and \eqref{dg2b1} to Poisson gauge: 
\bea
\delta_{g}&=&\delta_{g{\rm T}}  + \blue{(3-b_e){\cal H} V}\,,\label{dg1bp}
\\ \label{dg2bp}
\delta^{\tw}_{g} &=& \delta^{\tw}_{g{\rm T}} +\blue{(3-b_e){\cal H}V^{\tw} + \Big[ (b_e-3){\cal H}'+(b_e-3)(b_e-4){\cal H}^2 +b_e'{\cal H}\Big]\big( V\big)^2}
\notag \\
&&\blue{{}+ 2(3-b_e){\cal H} V\,\delta_{g{\rm T}} - 2V\,\delta_{g{\rm T}}'
+2(3-b_e){\cal H} V\,\Psi}\,,
\eea
{where the GR corrections in {magenta} scale as $(\cH^2/k^2)\,\delta_{\rm T}$ at first order, and as $(\cH^2/k^2)\,(\delta_{\rm T})^2$ or $(\cH^4/k^4)\,(\delta_{\rm T})^2$ at second order.}
For \eqref{dg2bp} we followed \cite{Bertacca:2014wga,  Jolicoeur:2017nyt,deWeerd:2019cae}, but we significantly  simplified their expressions, using the first-order Euler equation $V'+{\cal H}V=- \Psi$ and the {relation 
\be \label{vpsi}
 V=-{2f \over 3\Omega_m\cH}\, \Psi\,, 
\ee
which follows from the continuity equation, $\delta_{\rm T}'=-\nabla^2 V$, and the Poisson equation.}  We also included the evolution bias terms that are omitted in \cite{Umeh:2019jqg}.\\

\subsection{{Second-order metric and velocity potentials}} \label{ss-mvp}

{At second order,  the number density contrast has a GR correction in addition to a PNG correction, as shown in \eqref{dg2b1}.
 Unlike the  first-order case, the metric and velocity potentials at second order also have nonzero GR and PNG corrections:
\bea
\Psi^{\tw} &=& \Psi^{\tw}_{{\rm N}}+\blue{\Psi^{\tw}_{{\rm GR}}}+ \Psi^{\tw}_{{\rm nG}}\,, \\
\Phi^{\tw} &=& \Psi^{\tw}_{{\rm N}}+\blue{\Phi^{\tw}_{{\rm GR}}}+ \Psi^{\tw}_{{\rm nG}}\,, \\
V^{\tw} &=& V^{\tw}_{{\rm N}}+\blue{V^{\tw}_{{\rm GR}}}+ V^{\tw}_{{\rm nG}}\,,
\eea
where we note that 
\be \label{potnng}
\Phi^{\tw}_{{\rm N}}=\Psi^{\tw}_{{\rm N}} \quad \mbox{and} \quad \Phi^{\tw}_{{\rm nG}}=\Psi^{\tw}_{{\rm nG}}\,.
\ee
 The GR corrections are derived in  \citep{Villa:2015ppa} (which only considers modes $k<k_{\rm eq}$). Here we derive the PNG contributions, which include modes $k > k_{\rm eq}$.

The PNG corrections to metric and velocity potentials are linearly evolved, i.e., their nonlinearity is purely primordial, the same as in the case of the density contrast. They follow from  constraint and energy conservation equations applied to the linearly evolved PNG part of the matter density contrast, $\delta^{\tw}_{{\rm T,{nG}}}$. As we argued in deriving \eqref{alpha2}, $\delta^{\tw}_{{\rm T,{nG}}}$ obeys the linear Newtonian Poisson equation. The same applies to the linearly evolved $ \Psi^{\tw}_{{\rm nG}}$. From the Newtonian Poisson equation we find that
\bea
 \Psi^{\tw}_{{\rm nG}}(a,\k) &=& -{3\Omega_m(a) \cH(a)^2 \over 2 k^2}\, \delta^{\tw}_{{\rm T,{nG}}}(a,\k) 
 \notag\\ \label{ps2ng}
 &=& -{10\over3}\fnl\,\bigg[1+{2f(a) \over 3\Omega_m(a)} \bigg]^{-1} \, T(k)\,\big(\varphi_{\rm p} * \varphi_{\rm p}\big)(\k)\,,
\eea
where we used \eqref{alpha} and \eqref{alpha2}.

{By \eqref{ps2ng}, $\Psi^{\tw}_{{\rm nG}}$ grows as $(1+2f/3\Omega_m)^{-1}$, and thus
\be \label{p2ngp}
\Psi^{\tw \prime}_{{\rm nG}} =-{2f \over \big(3\Omega_m+2f \big)}\bigg({f' \over f}+\cH + 2{\cH' \over \cH } \bigg)
\, \Psi^{\tw}_{{\rm nG}}\,.
\ee}

The first-order linear equation \eqref{vpsi}, based on energy conservation and the Poisson equation, extends to second order for the linearly evolved PNG parts of the velocity and the potential. This determines the PNG part of the velocity:
\be \label{v2ng}
V^{\tw}_{{\rm nG}} = -{2f\over 3\Omega_m\cH }\,\Psi^{\tw}_{{\rm nG}}\,.
\ee
The linearly evolved PNG part of the second-order RSD term then follows as
\be \label{k2ng}
\partial_\|^2V^{\tw}_{{\rm nG}}(a,\k) = -2\fnl\,\cH(a) f(a)\mu^2\, {\cal M}(a,k) \,\big(\varphi_{\rm p} * \varphi_{\rm p} \big)(\k) \,,
\ee
where $\partial_{\parallel}=\bm{n}\cdot\bm{\nabla}$ and  $\mu=\hat{\k}\cdot\n$.
Finally, the first-order linear relation $\Phi=\Psi$ extends to second order for the linearly evolved PNG part of  $\Phi^{\tw}$, giving the second equality of \eqref{potnng}.}


\section{Local primordial non-Gaussianity in the relativistic bispectrum}\label{sec3}
\vspace*{0.5cm}
\subsection{Matter bispectrum}
~\\
{The primordial contribution of matter, independent of halo formation,  is given by the Newtonian approximation \eqref{alpha3},  corrected by the GR contribution in \eqref{dtgr2}:
\be
\delta^{\tw}_{{\rm T}} = \delta^{\tw}_{{\rm T,N}} + {\delta^{\tw}_{{\rm T,nG}}} + \blue{{20\over 3}\, \delta_{{\rm T}}\,{\hat\varphi_{{\rm in}}} -{5\over3}\, \xi^i\,\partial_i\,{\hat\varphi_{{\rm in}}}}\,. \label{dm2b}
\ee
The kernels in Fourier space corresponding to the GR terms in \eqref{dm2b} are:
\bea
\delta_{{\rm T}}\,{\hat\varphi_{\rm in}}~\rightarrow~- {\big(k_1^2+ k_2^2\big)\over 2 k_1^2 k_2^2}\,,\quad
 \xi^i\,\partial_i\,{\hat\varphi_{\rm in}} ~\rightarrow~ - {{\k}_1\cdot {\k}_2 \over k_1^2 k_2^2} \,.
\eea
Then the  tree-level matter bispectrum  $\big\langle \delta_{\rm T}\, \delta_{\rm T}\,\delta^{\tw}_{\rm T} \big \rangle$ at equal times is  given  by 
\begin{eqnarray}
B_m(\bm{k}_{1},\bm{k}_{2},\bm{k}_{3}) &=& \bigg\{ F_{2}(\bm{k}_{1},\bm{k}_{2})
+{2\fnl}{{\cal M}(k_3) \over {\cal M}(k_1) {\cal M}(k_2)}
\notag\\ &&{}~
-\blue{\big(3\Omega_m+2f\big) \cH^2 \, {\big[2\big(k_1^2+ k_2^2\big) - {\k}_1\cdot {\k}_2 \big]\over  2k_1^2 k_2^2}} \bigg\}
P(k_{1})P(k_{2}) + \mbox{2 cp}\,, \label{e1.11} 
\end{eqnarray} 
where  we  omit the time dependence for brevity, and `cp' denotes cyclic permutation. 
Here
$P\equiv P_{\rm T}$ is the linear matter power spectrum and}
\be
{{\cal M}(k_3) \over {\cal M}(k_1) {\cal M}(k_2)} = {3\over 10} \big(3\Omega_m+2f \big)
\cH^2\,{T(k_3) \over T(k_1) T(k_2)}\,{k_3^2 \over k_1^2 k_2^2}\,.
\ee 
The standard Newtonian result (see e.g. \cite{Tellarini:2015faa}) is modified in GR by the magenta terms in \eqref{e1.11}. 
For Gaussian initial conditions, the GR correction is suppressed by $\cH^2/k^2$ relative to the Newtonian approximation, but {\em in the non-Gaussian case, the GR correction is of the same order of magnitude as the local PNG term.}
\\

 

\subsection{{Observed number density}}\label{ss-ond}
~\\
The observed number density contrast is $\Delta_{g}+ \Delta_{g}^{(2)}/2$, which modifies the source  quantity $\delta_{g}+ \delta_{g}^{(2)}/2$ by RSD and other redshift space effects. It  can be split into {Newtonian, relativistic and non-Gaussian parts as follows. 
\begin{itemize}
\item
The {\bf first order} parts are:
\begin{eqnarray}
 \Delta_{g{\rm N}} &=& b_{10}\delta_{{\rm T,N}} - \frac{1}{\HH}  \partial_{\|}^2 V\,, \label{d1n} \\ 
 \Delta_{g\rm nG} &=& b_{01} \varphi_{\rm p}\,,
 \label{eq:linearNewtonian}
\\
\blue{\Delta_{g{\rm GR}}} &\blue{=}& \blue{\Big[ b_e   - 2 \mathcal{Q}  + \frac{2\left( \mathcal{Q}-1 \right)}{\chi \HH} - \frac{\HH'}{\HH^2} \Big]\big( \partial_{\|} V-  \Psi\big)}
\notag\\
&&{}\blue{+\,(2 \mathcal{Q} -1)  \Psi   + \frac{1}{\HH} \Psi' + (3- b_e )\HH V}\,.
\,\label{eq:linearGR}
\end{eqnarray}
Recall that $\delta_{\rm T}$, $V$ and $\Psi$ have no GR and no PNG corrections, by \eqref{dgro1} and \eqref{pvo1}.

\item
The {\bf second-order Newtonian} part of the observed number density contrast is formed from the density contrast and RSD terms and their couplings:
\begin{eqnarray} 
\Delta^{{{({2})}}}_{g{\rm N}}
  &=& \delta_{g{\rm T,N}}^{{{({2})}}}   - \frac{1}{\mathcal{H}}\partial_{\parallel}^{2}V^{{{({2})}}}_{\rm N}
  \notag\\
&&{}   - 2\frac{b_{10}}{\mathcal{H}}\bigg[\delta_{\rm T}\,\partial_{\parallel}^{2}V + \partial_{\parallel}V\,\partial_{\parallel}\delta_{\rm T}\bigg]
  + \frac{2}{\mathcal{H}^{2}}\bigg[\big(\partial_{\parallel}^{2}V\big)^{2} + \partial_{\parallel}V\,\partial_{\parallel}^{3}V\bigg]. \label{eq:SecondorderNewtonian}
 \eea
 
\item
  The {\bf second-order relativistic} part is  \cite{Jolicoeur:2017nyt,Jolicoeur:2017eyi}:\blue{
 \bea
\Delta_{g\rm GR}^{(2)}&=&\delta^{\tw}_{g{\rm T,GR}} 
- \frac{1}{\mathcal{H}}\partial_{\parallel}^{2}V_{\rm GR}^{(2)}+ \bigg[b_{e} - 2\mathcal{Q} +\frac{2(\mathcal{Q} -1)}{\chi\mathcal{H}}- \frac{\mathcal{H}'}{\mathcal{H}^{2}} \bigg]\left[\partial_{\parallel}V^{(2)}_{\rm N+GR} -\Phi^{(2)}_{\rm N+GR} \right] \notag \\
&&{}+2(\mathcal{Q}-1)\Psi^{(2)}_{\rm N+GR}  +\Phi^{(2)}_{\rm N+GR} + \frac{1}{\mathcal{H}}\Psi^{(2)\prime}_{\rm N+GR} +(3-b_e){\cal H}V^{\tw}_{\rm N+GR} \nonumber \\
\label{odg2}
&&{}+\mbox{very many terms quadratic in first-order  quantities,}
\eea}
where 
\be
V_{\rm N+GR}^{(2)}\equiv V_{\rm N}^{(2)}+\blue{V_{\rm GR}^{(2)}}\,, 
\ee
and similarly for the metric potentials.

The Newtonian parts of the metric potentials $\Psi^{\tw}, \Phi^{\tw}$ appear in the GR part of $\Delta_{g}^{(2)}$ because there is {\em no Newtonian projection effect involving these potentials}. For the velocity potential, the Newtonian part $V^{(2)}_{\rm N}$ is present only in the RSD term in \eqref{eq:SecondorderNewtonian}; the remaining velocity terms occur {\em only in the GR part} of $\Delta_{g}^{(2)}$ and therefore   $V^{(2)}_{\rm N}$ is included in the GR terms.

The  quadratic terms in \eqref{odg2} are given in full by \cite{Jolicoeur:2017nyt}. For convenience,  Appendix \ref{app1} presents all of the terms in  \eqref{odg2}, correcting some errors in \cite{Jolicoeur:2017nyt}. 

\item
 The {\bf second-order local PNG} part is 
 \begin{align}
\Delta_{g\rm nG}^{(2)}&=\delta^{\tw}_{g{\rm T,nG}} 
- \frac{1}{\mathcal{H}}\partial_{\parallel}^{2}V_{\rm nG}^{(2)}\notag \\
&{} 
+ \blue{\bigg[b_{e} - 2\mathcal{Q} +\frac{2(\mathcal{Q}-1)}{\chi\mathcal{H}}- \frac{\mathcal{H}'}{\mathcal{H}^{2}} \bigg]\!\!\left[\partial_{\parallel}V^{(2)}_{\rm nG} -\Psi^{(2)}_{\rm nG} \right]
 +(2\mathcal{Q}-1)\Psi^{(2)}_{\rm nG}   + \frac{1}{\mathcal{H}}\Psi^{(2)\prime}_{\rm nG} +(3-b_e){\cal H}V^{\tw}_{\rm nG}}
 \nonumber \\
\label{ngdg2}
&{}  - 2\frac{b_{01}}{\mathcal{H}}\bigg(\varphi_{\rm p}\,\partial_{\parallel}^{2}V + \partial_{\parallel}V\,\partial_{\parallel}\varphi_{\rm p}\bigg) \notag\\
&{} + \blue{b_{01} \Big(c_1 \Psi \varphi_{\rm p} + c_2 V \varphi_{\rm p}+c_3 \varphi_{\rm p} \partial_{\parallel}V + c_4 \Psi \partial_{\parallel}\varphi_{\rm p} \Big)}\,.
\end{align}
In this expression,  lines 1 and 2 contain the linearly evolved second-order terms  whose nonlinearity is purely primordial. Lines 3 and 4 contain the quadratic coupling terms. 

Line 1 is the Newtonian density + RSD  part, given by \eqref{dg2bn} and  \eqref{k2ng}. 

Line 2 arises from {\em GR projection terms that are absent in the Newtonian approximation:} these terms are given by \eqref{potnng}--\eqref{k2ng}. 

Line 3 arises from the first quadratic RSD term  in line 2 of \eqref{eq:SecondorderNewtonian}, given by the coupling of $\delta_{\rm T,nG}$ to velocity gradients.

Line 4 arises from the {\em coupling of $\delta_{\rm T,nG}$  to first-order GR projection terms.}  The  coefficients $c_I(a)$ are explicitly given below and in Appendix \ref{app2}.

{Apart from the $b_{02}$ term in $\delta^{\tw}_{g{\rm T,nG}}$, the Newtonian terms in \eqref{ngdg2} scale as $(\cH^2/k^2)\,(\delta_{\rm T})^2$  and dominate the  GR correction terms, which scale as ${\rm i}(\cH^3/k^3)\,(\delta_{\rm T})^2$ or $(\cH^4/k^4)\,(\delta_{\rm T})^2$.}

{In summary  the local PNG part at second order has the following origins:
\begin{itemize}
\item[\bf *] the primordial matter density contrast;
\item[\bf *]  the scale-dependent bias; 
\item[\bf *]  the linearly evolved second-order projection effects in velocity and metric potentials -- from RSD and from GR corrections;
\item[\bf *]  the coupling of first-order scale-dependent bias with first-order projection effects -- from RSD and from GR corrections. \\

\end{itemize} }

\end{itemize}
}

\subsection{Galaxy bispectrum}
~\\
At leading order the observed galaxy bispectrum is defined by \citep{Umeh:2016nuh}
\bea
 2\big \langle \Delta_{g}({\bm{k}_{1}}) \Delta_{g}({\bm{k}_{2}}) \Delta_{g}^{(2)}({\bm{k}_{3}}) \big \rangle + \mbox{2 cp} =(2\pi)^{3} B_{g}(\bm{k}_{1},\bm{k}_{2},\bm{k}_{3})\delta^{\mathrm{Dirac}}(\bm{k}_{1}+\bm{k}_{2}+\bm{k}_{3})\,, \label{e2.13_1} 
\eea
where here, and below, we  omit the time dependence for brevity and we assume equal-time correlations.
The bispectrum can be written in terms of Fourier kernels as
\bea \label{bisker}
 B_{g}(\bm{k}_{1},\bm{k}_{2},\bm{k}_{3}) = \mathcal{K}(\k_1)\, \mathcal{K}(\k_2)\, \mathcal{K}^{(2)}(\k_1,\k_2,\k_3)\,P(k_1)P(k_2)  + \mbox{2 cp},
\eea
where
\bea
\Delta_{g}(\k) &=&\mathcal{K}(\k) \, \delta_{\rm T}(\k)\,,\label{k1} \\
\Delta_{g}^{(2)}(\k_3) &=&\int {\ud\k_1 \over (2\pi)^3}\,\ud \k_2\, \delta^{\mathrm{Dirac}}(\bm{k}_{1}+\bm{k}_{2}-\bm{k}_{3})\, \mathcal{K}^{(2)}(\k_1,\k_2,\k_3)\, \delta_{\rm T}(\k_1)\,  \delta_{\rm T}(\k_2)  .\label{k2}
\eea

 In \citep{Jolicoeur:2017eyi}, the Newtonian and GR kernels   are presented, including all local relativistic effects, from projection, evolution and bias, but in the case of Gaussian initial conditions. Here we have updated these results and extended them to include the effects of local PNG. From Section \ref{ss-ond},  we find the following kernels.

\begin{itemize}
\item
At {\bf first order}, using \eqref{d1n}--\eqref{eq:linearGR} and \eqref{k1}:
\bea
\mathcal{K}_{\mathrm{N}}(\bm{k}_{a}) &=& b_{10}+f\mu_a^{2}\,,
\label{kn} \\ 
\blue{\mathcal{K}_{\mathrm{GR}}(\bm{k}_{a})} &\blue{=}& \blue{\mathrm{i}\,\mu_{a} \frac{\gamma_{1}}{k_{a}} + \frac{\gamma_{2}}{k_{a}^{2}}}\,,  \label{e2.21}\\
\mathcal{K}_{\rm nG}(\bm{k}_a) &=& \frac{b_{01}}{{\cal M}(k_a)}\,, \label{e2.14} 
\eea
where   $\mu_a=\hat{\k}_a\cdot\n$ and \blue{
\bea
\frac{\gamma_{1}}{\cH} &=& f\bigg[b_{e}-2\Q-\frac{2(1-\Q)}{\chi \cH} - \frac{\cH'}{\cH^{2}}\bigg], 
\label{gam1}\\
\frac{\gamma_{2}}{\cH^{2}} &=& f(3-b_{e})+\frac{3}{2}\Omega_{m}\bigg[2+b_{e}-f-4\Q-\frac{2(1-\Q)}{\chi \cH}-\frac{\cH'}{\cH^{2}}\bigg] . \label{e2.22}
\eea}

\item
The  {\bf second-order Newtonian part} follows from \eqref{eq:SecondorderNewtonian}  and \eqref{k2} (see e.g. \cite{Tellarini:2016sgp}): 
\bea
\mathcal{K}^{(2)}_{\mathrm{N}}(\bm{k}_{1}, \bm{k}_{2},\bm{k}_3) &=& b_{10}F_{2}(\bm{k}_{1}, \bm{k}_{2}) + b_{20} + f\mu_{3}^{2}G_{2}(\bm{k}_{1}, \bm{k}_{2}) + b_{s}S_{2}(\bm{k}_{1}, \bm{k}_{2}) 
 \label{e2.15}\\ &&{}
+b_{10}{f \big(\mu_1k_1+\mu_2k_2\big)\Big({\mu_1 \over k_1}+{\mu_2 \over k_2} \Big)} 
+  f^2{\mu_1\mu_2 \over k_1k_2}\big( \mu_1k_1+\mu_2k_2\big)^2 ,
\nonumber 
\eea 
where
\bea
G_{2}(\bm{k}_{1},\bm{k}_{2}) &=&{F'\over DD'}+ \Big({k_1\over k_2}+ {k_2\over k_1}\Big)\hat{\k}_1\cdot \hat{\k}_2 +\bigg(2-{F'\over DD'}\bigg)\big( \hat{\k}_1\cdot \hat{\k}_2\big)^2\,, \label{g2k}
\\
S_{2}(\bm{k}_{1}, \bm{k}_{2}) &=&\big( \hat{\k}_1\cdot \hat{\k}_2\big)^2- \frac{1}{3}\,. \label{e2.18}
\eea
Since we use the approximation $F/D^2=3/7$ in $F_2$,  we have $F'/(DD')=6/7$ in $G_2$.

\item
The {\bf second-order relativistic part} follows from \eqref{odg2} and \eqref{k2} (see \citep{Jolicoeur:2017eyi}, with some errors that are corrected here):\blue{ 
\begin{eqnarray}
\mathcal{K}^{(2)}_{\mathrm{GR}}(\bm{k}_{1}, \bm{k}_{2}, \bm{k}_{3}) &=& \frac{1}{k_{1}^{2}k_{2}^{2}}\Bigg\{\beta_{1} {+ E_{2}(\bm{k}_{1}, \bm{k}_{2},\bm{k}_3)\,\beta_2}  
\label{e2.23} \\
&& {} 
\qquad \quad + {\rm i}\bigg[\left(\mu_{1}k_{1} + \mu_{2}k_{2}\right)\beta_{3}  {+ {\mu_{3}k_{3}{\Big(\beta_4+E_{2}(\bm{k}_{1}, \bm{k}_{2},\bm{k}_3)\,\beta_5 \Big)}}}\bigg]   \nonumber \\
&& {} \qquad \quad   + \frac{k_{1}^2k_{2}^2}{k_{3}^2} \Big[{F_{2}(\bm{k}_{1}, \bm{k}_{2})}\,\beta_{6} + {G_{2}(\bm{k}_{1}, \bm{k}_{2})}\,\beta_{7} \Big] + \left(\mu_{1}k_{1}\mu_{2}k_{2}\right)\beta_{8}  
\nonumber \\
&& {} 
\qquad \quad +{{\mu_{3}^{2}k_{3}^{2}{\Big[\beta_9+ E_{2}(\bm{k}_{1}, \bm{k}_{2},\bm{k}_3)\,\beta_{10} \Big]}}} + \left(\bm{k}_{1}\cdot \bm{k}_{2}\right)\beta_{11} 
\nonumber \\
&& {}
 \qquad \quad 
+ \left(k_{1}^{2} + k_{2}^{2}\right)\beta_{12}+ \left(\mu_{1}^{2}k_{1}^{2} + \mu_{2}^{2}k_{2}^{2}\right)\beta_{13} 
\nonumber \\
&& {} 
\qquad \quad+ {\rm i}\bigg[\left(\mu_{1}k_{1}^{3} + \mu_{2}k_{2}^{3}\right)\beta_{14} + \left(\mu_{1}k_{1} + \mu_{2}k_{2}\right)\left(\bm{k}_{1} \cdot \bm{k}_{2}\right)\beta_{15}
\nonumber \\
&& {} 
\qquad \quad    + k_{1}k_{2}\left(\mu_{1}k_{2} + \mu_{2}k_{1}\right)\beta_{16}+ \left(\mu_{1}^{3}k_{1}^{3}+ \mu_{2}^{3}k_{2}^{3}\right)\beta_{17}
 \nonumber \\ \notag
&& {} 
\qquad \quad   + \mu_{1}\mu_{2}k_{1}k_{2}\left(\mu_{1}k_{1} + \mu_{2}k_{2}\right)\beta_{18}
+ \mu_{3}\frac{k_{1}^{2}k_{2}^{2}}{k_{3}}\,{G_{2}(\bm{k}_{1}, \bm{k}_{2})}\,\beta_{19}\bigg]\Bigg\}, 
\end{eqnarray}}
where
\begin{equation}
E_{2}(\bm{k}_{1}, \bm{k}_{2},\bm{k}_3)=\frac{k_{1}^{2}k_{2}^{2}}{k_{3}^{4}}
\bigg[3+2\bigg(\frac{k_{1}}{k_{2}} + \frac{k_{2}}{k_{1}}\bigg) \hat{\k}_1\cdot \hat{\k}_2 + \big( \hat{\k}_1\cdot \hat{\k}_2\big)^2\bigg]\,. 
\end{equation}
The kernel \eqref{e2.23} is derived from the many terms in $\Delta_g^{(2)}(\x)$, as given in  \cite{Bertacca:2014dra, Bertacca:2014hwa} (we neglect the integrated terms). For convenience, in Tables \ref{tabc1} and \ref{tabc2},  Appendix \ref{app1},  we summarise  which terms in $\Delta_g^{(2)}(\x)$ contribute to which of the terms in  \eqref{e2.23}. The time-dependent functions $\beta_{I}$ are also given in Appendix \ref{app1}.
 
\item
The {\bf second-order local PNG part}  follows from \eqref{ngdg2}:
\bea
\mathcal{K}^{(2)}_{\mathrm{nG}}(\bm{k}_{1},\bm{k}_{2},\bm{k}_{3}) &=& {2}\,f_{\mathrm{NL}}\big(b_{10}+f\mu_{3}^{2}\big)\frac{{\cal M}_{3}}{{\cal M}_{1}{\cal M}_{2}} 
+ fb_{01}\big(\mu_1k_1+\mu_2k_2\big) \bigg({\mu_1 \over k_1{\cal M}_{2}}+{\mu_2 \over k_2 {\cal M}_{1}} \bigg)
\notag \label{e2.16}
\\  && {}
+ b_{n}{N}_{2}(\bm{k}_{1},\bm{k}_{2})  + {b_{11}\over 2}\bigg(\frac{1}{{\cal M}_{1}} + \frac{1}{{\cal M}_{2}}\bigg) 
+\frac{b_{02}}{{\cal M}_{1}{\cal M}_{2}} 
\notag \\ & & 
\blue{{}+\frac{{\cal M}_3}{{\cal M}_{1}{\cal M}_{2}}\bigg( {\Upsilon_{1} \over k_3^2} + \mathrm{i}\,{\mu_{3} \over k_3}\,\Upsilon_{2} \bigg) 
+{\Upsilon_{3}} \bigg(\frac{1}{k_{1}^{2}{\cal M}_{2}} + \frac{1}{k_{2}^{2}{\cal M}_{1}}\bigg) }
\nonumber \\
 && 
\blue{{} + \mathrm{i}\,\bigg[{\Upsilon_{4}}\bigg(\frac{\mu_{1}k_{1}}{k_2^2{\cal M}_{1}} + \frac{\mu_{2}k_{2}}{k_1^2{\cal M}_{2}}\bigg)  
 +{\Upsilon_{5}} \bigg(\frac{\mu_{1}}{k_1{\cal M}_{2}} + \frac{\mu_{2}}{k_2{\cal M}_{1}}\bigg) \bigg] ,} \label{e2.35} 
\eea 
where ${\cal M}_a \equiv {\cal M}(k_a)$ and
\be
N_{2}(\bm{k}_{1},\bm{k}_{2}) = {1\over 2} \bigg({k_1 \over k_2{\cal M}_1 }+ {k_2 \over k_1{\cal M}_2 } \bigg) \hat{\k}_1\cdot \hat{\k}_2 . \label{n2k}
\ee

In the first line of \eqref{e2.16}, the first term is a sum of the  matter density term in line 2 of \eqref{dg2bn} and the linearly evolved PNG part of the second-order RSD term [line 1 of  \eqref{ngdg2}]. The second term is the quadratic RSD term from line 3 of \eqref{ngdg2}. 

The second line gives the scale-dependent bias contribution from \eqref{dg2bn}. The first two lines recover the Newtonian approximation (see \cite{Tellarini:2015faa}). 

Lines 3 and 4 in {magenta} are the PNG contributions that arise from relativistic projection effects, as explained in Section \ref{ss-ond}. 
These projection terms in the non-Gaussian kernel involve new time-dependent functions $\Upsilon_I$, which are given in Appendix \ref{app2}.
{The terms in $\Delta_g^{(2)}(\x)$ corresponding to those in \eqref{e2.35}, lines 3 and 4, are summarised in Table \ref{tabc3}, Appendix \ref{app2}.}

{The Newtonian terms scale as $(\cH^2/k^2)\,(\delta_{\rm T})^2$ except for the $b_{02}$ term which scales as $(\cH^4/k^4)\,(\delta_{\rm T})^2$. The relativistic $\Upsilon_1, \Upsilon_3$ terms scale as $(\cH^4/k^4)\,(\delta_{\rm T})^2$, while the $\Upsilon_2, \Upsilon_4, \Upsilon_5$ terms are  ${\cal O}(\cH^3/k^3)$.}

Note that $\Upsilon_1,\Upsilon_2$ are proportional to $\fnl$, and $\Upsilon_3,\Upsilon_4,\Upsilon_5$ are  proportional to $b_{01}$ (which itself is proportional to $\fnl$).

For Gaussian initial conditions, $\mathcal{K}^{(2)}_{\mathrm{nG}}$ vanishes: 
\be
\fnl=0~~\Rightarrow~~b_{01}=b_n=b_{11}=b_{02}=\Upsilon_I=0 ~~\Rightarrow~~ \mathcal{K}^{(2)}_{\mathrm{nG}}(\bm{k}_{1}, \bm{k}_{2}, \bm{k}_{3})=0\,.
\ee

\end{itemize}

\subsection{Numerical examples}

{The GR corrections to the Newtonian bispectrum, for both Gaussian and local PNG cases, are sensitive to the following astrophysical parameters of the tracer: Gaussian bias $b_{10}$, PNG bias $b_{01}$,  and magnification bias ${\cal Q}$, together with  their first derivatives in time and luminosity; evolution bias $b_e$ and its first  time derivative. This can be seen from the kernels presented above,  with the details given in Appendices \ref{app1} and \ref{app2}.} 

In order to illustrate the GR corrections, we need to use physically self-consistent values for these parameters, as well as for the second-order Newtonian clustering bias parameters $b_{20}$ and $b_s$.  
For a Stage IV H$\alpha$ spectroscopic survey, similar to Euclid,  we use \cite{Maartens:2019yhx} for the  clustering biases, evolution bias and magnification bias. {We neglect the luminosity derivatives of first-order clustering bias and magnification bias. For the PNG biases $b_{11}, b_n, b_{02}$ we use \eqref{b11}--\eqref{b02}.}

We start by showing the contribution of GR corrections to
the monopole of the reduced   bispectrum,
\be \label{redb}
Q^{00}_g(k_1,k_2,k_3) ={ B^{00}_g(k_1,k_2,k_3) \over P(k_1)P(k_2)+ P(k_3)P(k_1)+ P(k_2)P(k_3) }\,,
\ee
where \cite{deWeerd:2019cae}
\be \label{monob}
B^{\ell m}_g(k_1,k_2,k_3)= \int_0^{2\pi} \ud\phi \int_{-1}^1 \ud\mu_1 B_g(k_1,k_2,k_3,\mu_1,\phi)\,Y^*_{\ell m}(\mu_1,\phi) \,.
\ee
Here $\phi,\mu_1$ determine the orientation of the triangle relative to the line of sight. 
Figure \ref{qmono} shows the monopole
for squeezed configurations. We use fixed equal sides 
$k_1=k_2=0.1\,h$/Mpc and varying long mode $k_3<k_1=k_2$. The isosceles triangle is increasingly squeezed as $k_3$ decreases. 
The left panel shows the Newtonian approximation (dash-dot lines) and the right panel shows the monopole without the GR bias correction \eqref{dg2b}. 

\begin{figure}[!h]
\centering
\includegraphics[width=0.49\textwidth]{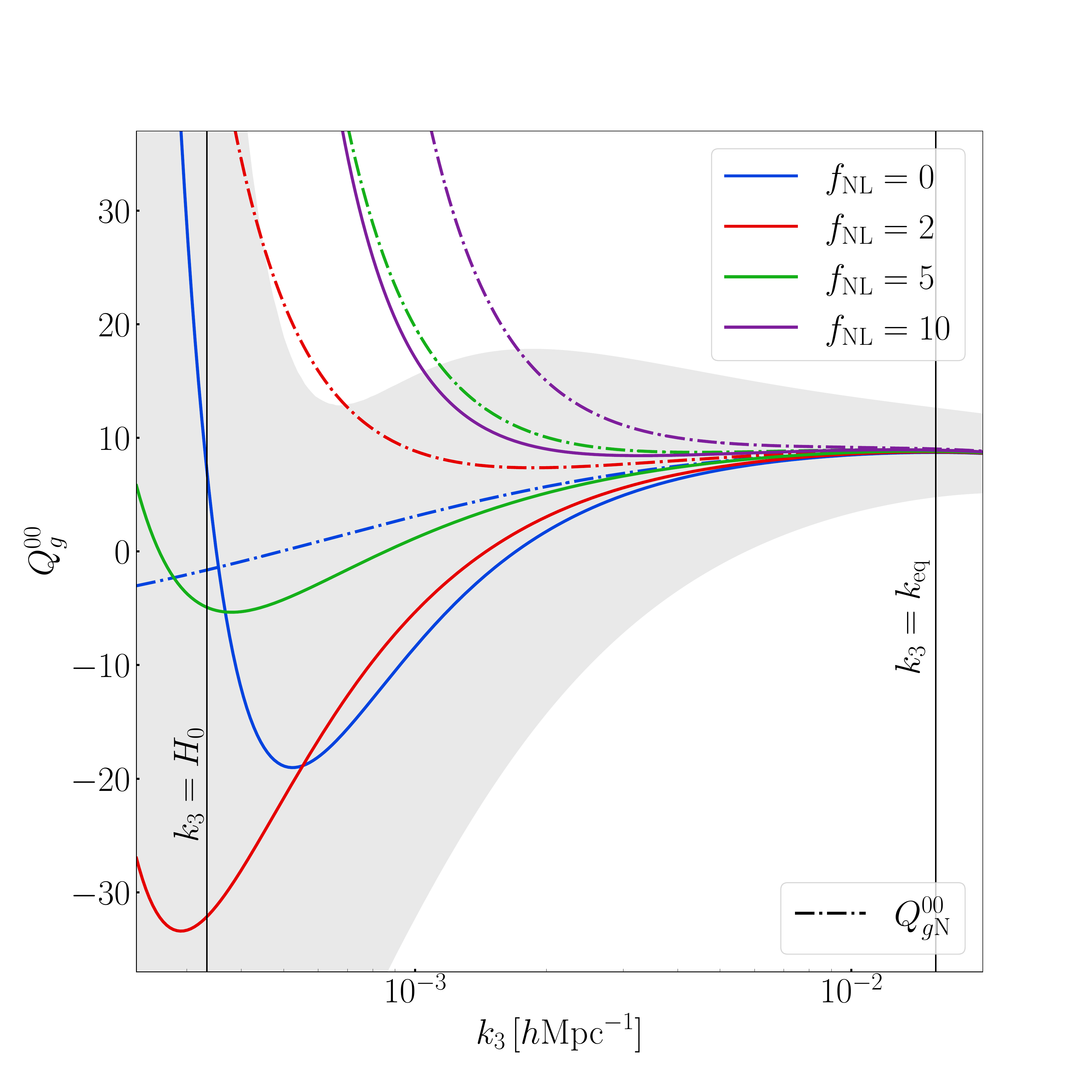}
\includegraphics[width=0.49\textwidth]{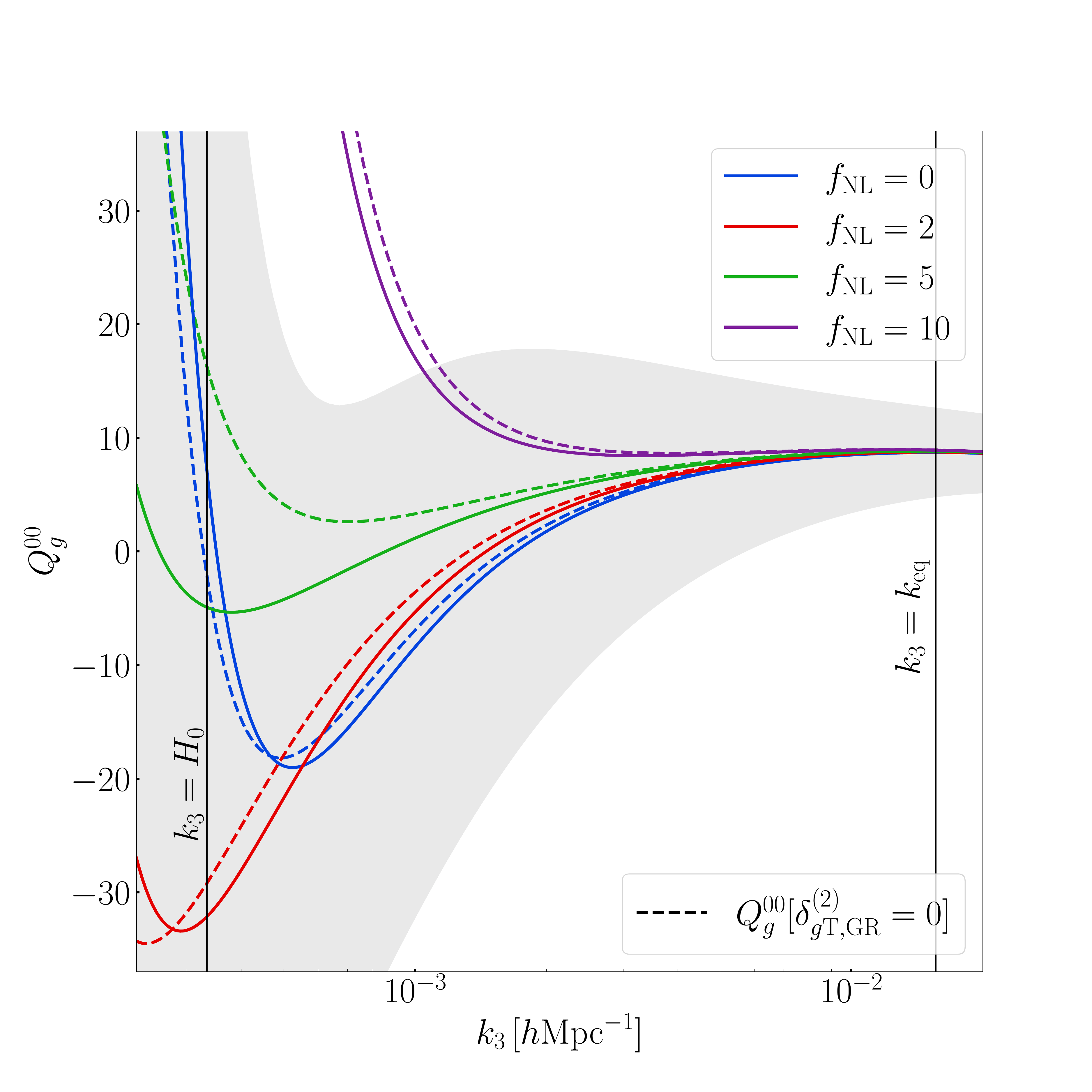}
\caption{ {Monopole of the reduced bispectrum for a Stage IV H$\alpha$ survey at $z=1$,  for various $\fnl$, with $k_1=k_2=0.1\,h$/Mpc. 
Shading indicates the 1$\sigma$ uncertainty (neglecting shot noise) for the $\fnl=0$ case (solid blue curve).
{\em Left:} Comparing the full relativistic monopole to the Newtonian approximation (dash-dot curves).   {\em Right:} Comparing the full relativistic monopole to the monopole without the GR correction to second-order  galaxy bias, \eqref{dtgr2} (dashed curves). }}
\label{qmono}
\end{figure}

The shading in Figure \ref{qmono} is defined by the {cosmic variance limited} error $\sigma_{B}$ on the $\fnl=0$ monopole, given by \cite{Gagrani:2016rfy}: 
\bea
{\big(\sigma_{B}\big)^2={{\cal V}^{\,\rm com} \over \pi k_1k_2 k_3\, \Delta k}\,
\int \ud \mu_1\, \ud \phi \,P_g(k_1,\mu_1) \,P_g(k_2,\mu_2)\, P_g(k_3,\mu_3)}
\,,\label{sigb}
\eea
where the galaxy power spectrum, from \eqref{kn}--\eqref{e2.14}, is
\bea
{P_g(k_a,\mu_a)=  \left| b_{10}
+f\mu_a^2+ {\gamma_2 \over k_a^2} +{\rm i}\,\mu_a{\gamma_1 \over k_a}\right|^2P(k_a)}\,. \label{pg0}
\eea
In \eqref{sigb}, ${\cal V}^{\,\rm com}$ is the comoving volume of the redshift bin, $\Delta k$ is chosen as the fundamental mode, $2\pi ({\cal V}^{\,\rm com})^{-1/3}$, $k_1=k_2=0.1\,h$/Mpc, and \cite{Clarkson:2018dwn} $\mu_2=\mu_1\cos\theta_{12}+\sqrt{1-\mu_1^2}\sin\theta_{12}\,\cos\phi$, $\mu_3=-(k_1\mu_1+k_2\mu_2)/k_3$. Here $\theta_{12}$ is the {tail-to-tail} angle between $\k_1$ and $\k_2$, so that the squeezed limit is $\theta_{12}=\pi$.

\begin{figure}[!h]
	\centering
	\includegraphics[width=\textwidth]{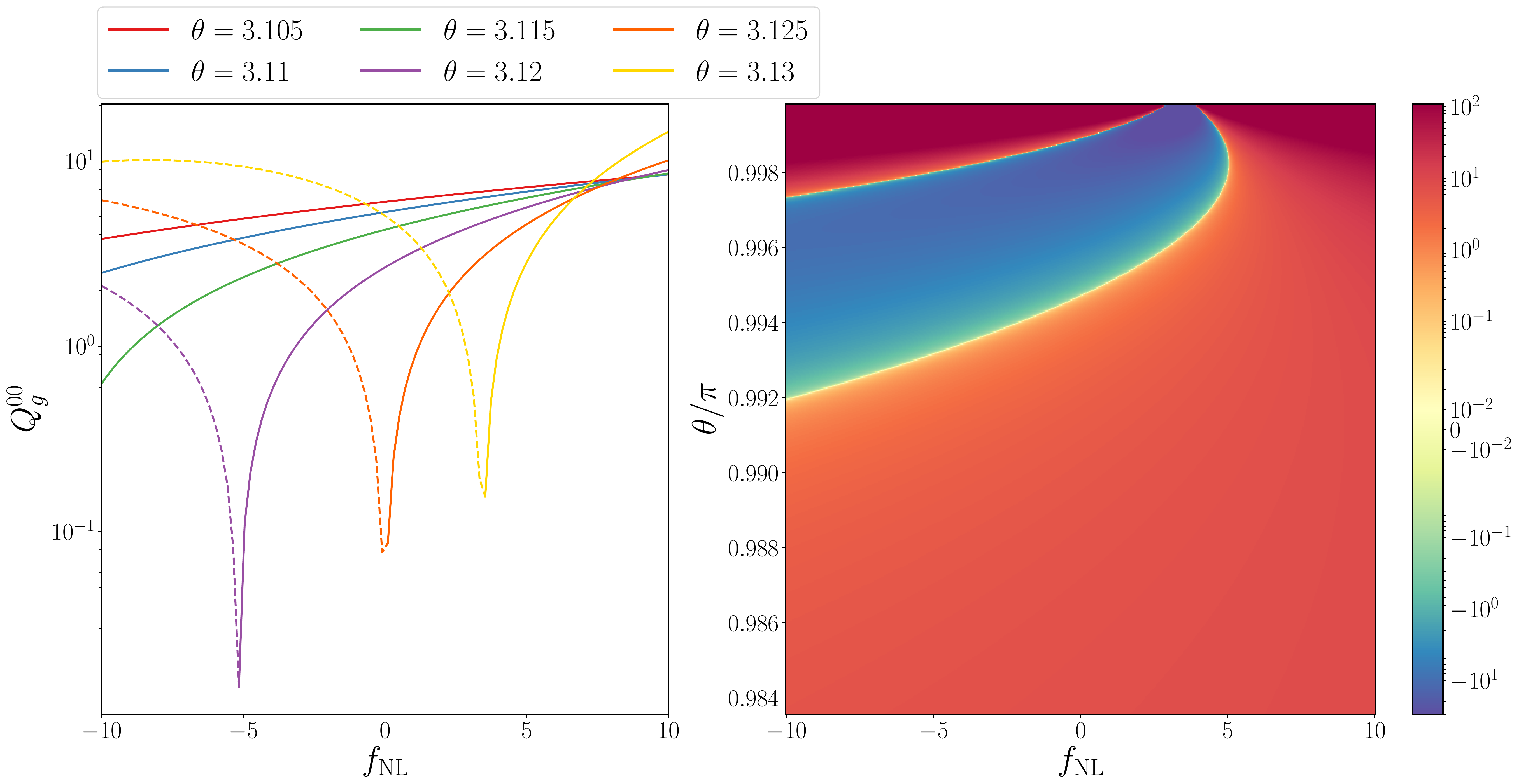}
\caption{Monopole of reduced bispectrum for isosceles triangles, as in Figure \ref{qmono}. {\em Left:} As a function of $\fnl$, for various values of $\theta\equiv\theta_{12}$, where $\theta=\pi$ is the squeezed limit. Dashed curves indicate negative values. {\em Right:} 2D colour map  as a function of $\fnl$ and $\theta/\pi$.}\label{rb2d}
\end{figure}
\begin{figure}[!h]
\centering
\includegraphics[width=\textwidth]{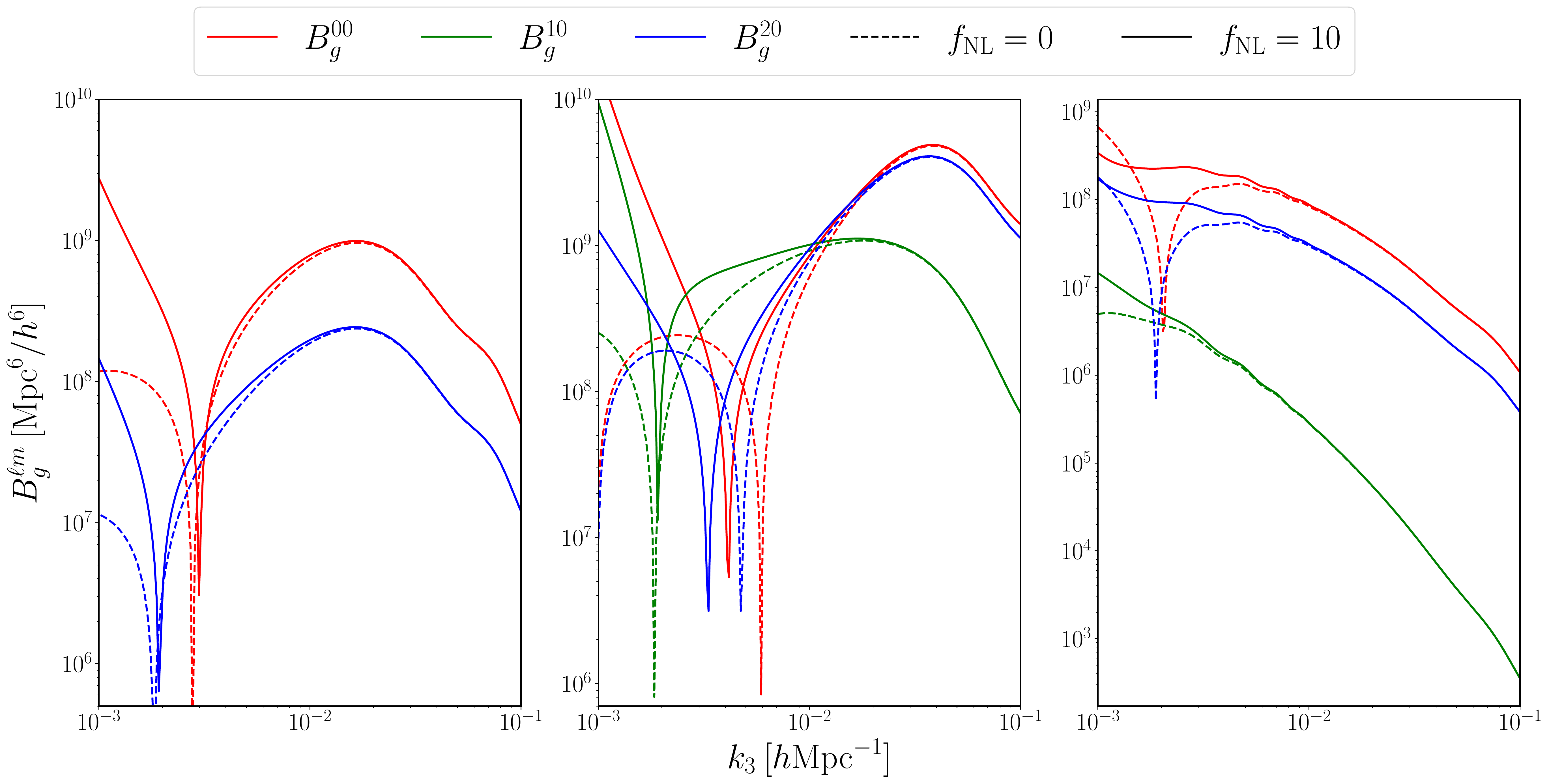}
\caption{First few nonzero multipoles for fixed triangle shape as a function of $k_3$, with {$\fnl=10$ (solid) and $\fnl=0$} (dashed).
{\em Left:} Equilateral configuration, $k_1=k_2=k_3$. {\em Middle:} Flattened configuration, $k_1=k_2\approx k_3/2$, with $\theta_{12}=2\degree$. {\em Right:} Squeezed configuration with $\theta_{12}=178\degree$ and $k_1=k_2=k_3/(2\sin \theta_{12})\approx 14\, k_3$.}
\label{eqfl}
\end{figure}

The effect of $\fnl$ is strongest in the monopole and competes with the GR contribution on ultra-large scales, since they both affect the Newtonian Gaussian bispectrum at ${\cal O}(\mathcal{H}^{2} / k^{2})$. We see this in Figure \ref{qmono} left panel, which shows the monopole of the reduced bispectrum for an increasingly squeezed isosceles triangle. 
In the Gaussian case (blue) we see that the Newtonian reduced monopole (dot-dash blue) becomes negative when the long mode is close to the Hubble scale, due to the effects of second-order galaxy bias.
The Gaussian GR correction to the Newtonian approximation is negative for super-equality long modes until close to the Hubble scale (this was pointed out in  \cite{Jolicoeur:2018blf}). GR effects drive the reduced monopole (solid blue) below zero for $H_0 \lesssim k_3\lesssim 0.002\,h$/Mpc  (the locations of the zero-crossings  are dependent on the Gaussian bias parameters, evolution bias and magnification bias). 

As $\fnl$ is increased above zero, the amplitude of the Newtonian reduced monopole (dot-dash curves) increases monotonically. When GR effects are taken into account, the reduced monopole is pushed upwards, but remains negative on observable scales for $\fnl \lesssim 5$, until it becomes always positive for $\fnl>5$   -- the precise turnaround value of $\fnl$ depends on astrophysical parameters. This means that for  $\fnl \lesssim 5$, 
local PNG {\em decreases} the amplitude of the reduced monopole on observable scales, in contrast to the Newtonian approximation. Comparing the green solid and blue dot-dash curves shows that {\em the Newtonian approximation is very close to the true reduced monopole with $\fnl \sim 5$}. For a universe with $\fnl \sim 5$, a Newtonian analysis of the squeezed bispectrum would conclude that the primordial universe is Gaussian. Similarly, a universe with $\fnl \sim 10$ would appear to have $\fnl \sim 5$ in a Newtonian approximation.

The GR contribution to the monopole is made up of: ${\cal O}(\mathcal{H}^{2} / k^{2})$ Gaussian projection terms, ${\cal O}(\mathcal{H}^{2} / k^{2})$  second-order galaxy bias correction (the same for Gaussian and PNG cases) and ${\cal O}(\mathcal{H}^{4} / k^{4})$ second-order local PNG contributions from GR projection effects.  The last contribution is effectively negligible on observable scales.
In the right panel of Figure \ref{qmono} we show that the GR bias correction is dominated by the Gaussian GR projection terms:  the effect of removing the GR correction to second-order galaxy bias is small. Note that the GR bias correction has a similar effect to a small negative value of $\fnl$.

In Figure \ref{rb2d}  we include  negative $\fnl$ and explore how  local PNG changes the monopole of the reduced bispectrum as we approach the squeezed limit, $\theta_{12}\to\pi$. For $\fnl\geq0$, the results provide a different perspective on Figure \ref{qmono} left panel. For negative $\fnl$, local PNG and GR effects act together to drive the monopole negative, so that the zero-crossing of the monopole occurs for smaller $\theta_{12}$, equivalently larger $k_3$. 

Figure \ref{eqfl} shows the effect of $\fnl$ on
the first three multipoles of the relativistic galaxy bispectrum, also including equilateral and flattened triangle shapes.
 In general, the Newtonian RSD effect induces only even multipoles, while the GR corrections modify the even multipoles and induce new odd multipoles. We show here the $m=0$  dipole (absent without GR corrections) and quadrupole (mainly Newtonian), compared to the monopole. 
 
 For the equilateral shape (left panel), the dipole vanishes exactly in the Gaussian case  \cite{Clarkson:2018dwn,deWeerd:2019cae} and nonzero $\fnl$ does not changes this result. The effect of $\fnl$ on the quadrupole is very similar to the case of the monopole. 
 
 For the flattened shape (middle panel), the dipole is the dominant part of the bispectrum for $0.002 \lesssim k_3/(h {\rm Mpc}^{-1})\lesssim 0.01$, and we see that $\fnl>0$ increases this effect further. The dipole $B_g^{1m}$ is purely relativistic: it vanishes in the Newtonian approximation  \cite{Clarkson:2018dwn,Maartens:2019yhx,Jolicoeur:2020eup}. 
 
 Finally, in the squeezed case (right panel), the effect on the monopole of  $\fnl=10$ is consistent with Figure \ref{qmono}. The quadrupole has a similar behaviour, and dominates the dipole. It is interesting that the three multipoles are approximately equal at scales near $k= 0.002\,h$/Mpc. Once again, this value is sensitive to astrophysical parameters.


\section{Conclusions}\label{sec4}

Upcoming galaxy surveys  and 21cm intensity mapping surveys will deliver high-precision cosmological measurements and constraints, based on a combination of the power spectrum and bispectrum. This advance demands a commensurate advance in theoretical precision. Here we contribute to the development of theoretical precision by deriving for the first time the local relativistic corrections to the tree-level redshift-space bispectrum in the presence of local primordial non-Gaussianity (PNG).

At first order in perturbations, there are no relativistic corrections to the comoving matter and galaxy density contrasts -- and therefore no correction to the galaxy clustering bias relation.  There are also no relativistic corrections to the velocity and metric potentials.  Consequently, there is no relativistic contribution to local PNG. The only relativistic correction is to the Newtonian projection effect, i.e. standard redshift-space distortions (RSD). 

At second-order, relativistic corrections go beyond projection effects to alter the galaxy bias relation and local PNG in the galaxy bispectrum. In summary, there are: 
\begin{itemize}
\item
relativistic projection corrections to the Newtonian RSD at first and second order;
\item
relativistic corrections to the Newtonian bias model in the comoving frame at second order;
\item
 second-order relativistic projection  corrections to the local PNG carried by Newtonian RSD -- from a coupling of first-order scale-dependent bias to first-order relativistic projection effects,  and from the linearly evolved local PNG in second-order velocity and metric potentials. 
 \end{itemize}

Our previous work \cite{Umeh:2016nuh,Jolicoeur:2017nyt, Jolicoeur:2017eyi, Jolicoeur:2018blf,Clarkson:2018dwn,Maartens:2019yhx,deWeerd:2019cae, Jolicoeur:2020eup} presented local (non-integrated) relativistic effects in the case of primordial Gaussianity and without the relativistic correction to galaxy bias. We have made corrections to these earlier results. In addition,  we have presented for the first time the galaxy bispectrum with relativistic corrections to galaxy clustering bias and new local PNG contributions that are encoded in relativistic projection effects.
Our main results are given  in Fourier space in \eqref{e2.15}--\eqref{e2.16}, with further details in Appendices \ref{app1} and \ref{app2}.

In Figures \ref{qmono} and \ref{rb2d}  we show examples of the squeezed monopole of the  reduced relativistic bispectrum for a Stage IV H$\alpha$ survey similar to Euclid, using physical models for the astrophysical parameters (clustering biases, evolution bias, magnification bias). These figures reveal various interesting relativistic features. In particular, they show  the bias in the estimate of $\fnl$ from using a Newtonian analysis. This bias is given by
\be
f_{\rm NL}^{\rm Newt} = \fnl + \Delta \fnl\,.
\ee
For the Stage IV survey at $z=1$, the bias can be roughly estimated by eye as $\Delta \fnl\sim 5$, for the long mode above the equality scale. Although the precise level of bias is sensitive to astrophysical parameters and redshift, the point is that next-generation precision demands that relativistic corrections are included in the bispectrum.

\bro{In common with nearly all work on the Fourier-space bispectrum with RSD and PNG, we implicitly make a flat-sky assumption, based on the fixed global direction $\bm n$.  
As a consequence, wide-angle correlations are not included, so that the flat-sky analysis loses accuracy as $\theta$ increases, where $\theta$ is the maximum opening angle to 
the three-point correlations at the given redshift. This leads to a systematic bias in the separation of observational effects from the PNG signal, and therefore in the best-fit value  of $f_{\rm NL}$.
 Including wide-angle effects  is a key target for future work. 
Corrections to the global flat-sky analysis of  the Fourier bispectrum can be made by using a local or `moving' line of sight  \cite{Scoccimarro:2015bla,Sugiyama:2018yzo,Shirasaki:2020vkk}.
However, corrections of this type are approximate and do not incorporate all the wide-angle effects. Ultimately, one needs to use the full-sky 3-point correlation function or the full-sky angular bispectrum (see e.g. \cite{Kehagias:2015tda, DiDio:2016gpd, DiDio:2018unb, Durrer:2020orn}) to properly include all wide-angle correlations. A major problem is that both of these alternatives are computationally more intensive. }

\vspace{2cm}
\noindent {\bf Acknowledgements}\\
\noindent
We thank Kazuya Koyama for very helpful discussions and Alexandre Barreira for a useful comment.
RM and SJ are supported by  the South African Radio Astronomy Observatory (SARAO) and the National Research Foundation (Grant No. 75415). 
RM and OU are supported by the UK Science \& Technology Facilities Council (STFC) Consolidated Grant ST/S000550/1. 
CC is supported by STFC Consolidated Grant ST/P000592/1.

\clearpage
\appendix
\section{$\beta_I$ functions in \eqref{e2.23}} \label{app1}
\begin{eqnarray} 
\frac{\beta_{1}}{\mathcal{H}^{4}} &=& \frac{9}{4}\Omega_{m}^{2}\Bigg[6-2f\bigg(2b_{e}-4\Q-\frac{4(1-\Q)}{\chi\cH}-\frac{2\cH'}{\cH^{2}}\bigg)-\frac{2f'}{\cH}+b_{e}^{2}+5b_{e}-8b_{e}\Q + 4\Q + 16\Q^{2} \nonumber\\ 
&&{} \qquad - 16\frac{\p \Q}{\p \ln{{L}}} - 8\frac{\Q'}{\cH} + \frac{b_{e}'}{\cH}+\frac{2}{\chi^{2}\cH^{2}}\bigg(1-\Q+2\Q^{2}-2\frac{\p \Q}{\p \ln{{L}}}\bigg)
 \nonumber\\ 
&&{} \qquad  - \frac{2}{\chi\cH}\bigg(3+2b_{e}-2b_{e}\Q-3\Q +8\Q^{2}-\frac{3\cH'}{\cH^{2}}(1-\Q) -8\frac{\p \Q}{\p \ln{{L}}} - 2\frac{\Q'}{\cH}\bigg) \nonumber \\
&&{} \qquad + \frac{\cH'}{\cH^{2}}\bigg(-7-2b_{e}+8\Q+\frac{3\cH'}{\cH^{2}}\bigg) - \frac{\cH''}{\cH^{3}}\Bigg] \nonumber \\
&&{} +\frac{3}{2}\Omega_{m}f\Bigg[{5}-2f(4-b_{e})+\frac{2f'}{\cH}+{2b_{e}\bigg(5+\frac{2(1-\Q)}{\chi\cH}\bigg)}-\frac{2b_{e}'}{\cH} -2b_{e}^{2} + 8b_{e}\Q - 28\Q \nonumber \\
&&{} \qquad \qquad - \frac{14(1-\Q)}{\chi\cH}-\frac{3\cH'}{\cH^{2}} +4\bigg(2-\frac{1}{\chi\cH}\bigg)\frac{\Q'}{\cH}\Bigg] \nonumber \\
&&{} +\frac{3}{2}\Omega_{m}f^{2}\bigg[-2+2f-b_{e}+4\Q+\frac{2(1-\Q)}{\chi\cH}+\frac{3\cH'}{\cH^{2}}\bigg] \nonumber \\
&&{} +f^{2}\bigg[12-7b_{e}+b_{e}^{2}+\frac{b_{e}'}{\cH}+(b_{e}-3)\frac{\cH'}{\cH^{2}}\bigg] - \frac{3}{2}\Omega_{m}\frac{f'}{\cH} 
\\&& \nonumber\\
\frac{\beta_{2}}{\cH^{4}} &=& \frac{9}{2}\Omega_{m}^{2}\bigg[-1+b_{e}-2\Q-\frac{2(1-\Q)}{\chi\cH}-\frac{\cH'}{\cH^{2}}\bigg] + 3\Omega_{m}f\bigg[{-1}+2f{-b_{e}}+4\Q+\frac{2(1-\Q)}{\chi\cH}+\frac{3\cH'}{\cH^{2}}\bigg] \nonumber \\
&&{} +3\Omega_{m}f^{2}\bigg[-1+b_{e}-2\Q-\frac{2(1-\Q)}{\chi\cH}-\frac{\cH'}{\cH^{2}}\bigg] + 3\Omega_{m}\frac{f'}{\cH}\; 
\\&& \nonumber\\
\frac{\beta_{3}}{\mathcal{H}^{3}} &=& \frac{9}{4}\Omega_{m}^{2}(f-2+2\Q) 
 + \frac{3}{2}\Omega_{m}f\Bigg[-2 -f\bigg(-3+f+2b_{e}-3\Q-\frac{4(1-\Q)}{\chi\cH}-\frac{2\cH'}{\cH^{2}}\bigg)-\frac{f'}{\cH}\nonumber \\
&&{} \qquad \qquad +{3}b_{e}+b_{e}^{2}-6b_{e}\Q{+4}\Q +8\Q^{2}-8\frac{\p \Q}{\p \ln{{L}}} -6\frac{\Q'}{\cH} +\frac{b_{e}'}{\cH}  \\
&&{} \qquad \qquad +\frac{2}{\chi^{2}\cH^{2}}\bigg(1-\Q+2\Q^{2}-2\frac{\p \Q}{\p \ln{{L}}}\bigg) + \frac{2}{\chi \cH}\bigg({-1} -2b_{e}+2b_{e}\Q{+}\Q-6\Q^{2}  \nonumber \\
&&{} \qquad \qquad +\frac{3\cH'}{\cH^{2}}(1-\Q) +6\frac{\p \Q}{\p \ln{{L}}} + 2\frac{\Q'}{\cH}\bigg) -\frac{\cH'}{\cH^{2}}\bigg({3}+2b_{e}-6\Q-\frac{3\cH'}{\cH^{2}}\bigg) - \frac{\cH''}{\cH^{3}}\Bigg] \nonumber \\
&&{} {+} f^{2}\Bigg[-3+2b_{e}\bigg(2+\frac{(1-\Q)}{\chi\cH}\bigg)-b_{e}^{2}+2b_{e}\Q -6\Q-\frac{b_{e}'}{\cH}-\frac{6(1-\Q)}{\chi\cH}
 +2\bigg(1-\frac{1}{\chi\cH}\bigg)\frac{\Q'}{\cH}\Bigg]  \notag
\\&& \nonumber\\
\frac{\beta_{4}}{\cH^{3}} &=& \frac{9}{2}\Omega_{m}f\bigg[-b_{e}+2\Q+\frac{2(1-\Q)}{\chi \cH}+\frac{\cH'}{\cH^{2}}\bigg] 
\\ \nonumber\\
{{\frac{\beta_5}{\cH^{3}}}} &=& 3\Omega_{m}f\bigg[b_{e}-2\Q-\frac{2(1-\Q)}{\chi\cH}-\frac{\cH'}{\cH^{2}}\bigg]
\end{eqnarray}

\begin{eqnarray}
\frac{\beta_6}{\cH^{2}} &=& \frac{3}{2}\Omega_{m}\Bigg[2-2f+b_{e}-4\Q-\frac{2(1-\Q)}{\chi\cH}-\frac{\cH'}{\cH^{2}}\Bigg]  \\ &&\nonumber\\
\frac{\beta_7}{\cH^{2}} &=& f(3-b_{e}) \\ \nonumber \\
\frac{\beta_8}{\cH^{2}} &=& {3\Omega_{m}f(2-f-2\Q)} + f^{2}\Bigg[4+b_{e}-b_{e}^{2}+4b_{e}\Q-{6}\Q-4\Q^{2}+4\frac{\p \Q}{\p \ln{{L}}} + 4\frac{\Q'}{\cH} - \frac{b_{e}'}{\cH}  \nonumber \\\nonumber\\
&&{} - \frac{2}{\chi^{2}\cH^{2}}\bigg(1-\Q+2\Q^{2}-2\frac{\p \Q}{\p \ln{{L}}}\bigg) - \frac{2}{\chi\cH}\bigg(3-2b_{e}+2b_{e}\Q-\Q-4\Q^{2}+\frac{3\cH'}{\cH^{2}}(1-\Q) \nonumber\\
&&{} + 4\frac{\p \Q}{\p \ln{{L}}} + 2\frac{\Q'}{\cH}\bigg) - \frac{\cH'}{\cH^{2}}\bigg(3-2b_{e}+{4}\Q+\frac{3\cH'}{\cH^{2}}\bigg) + \frac{\cH''}{\cH^{3}}\Bigg] \\ \nonumber \\
\frac{\beta_{9}}{\cH^{2}} &=& -\frac{9}{2}\Omega_{m}f \\ \nonumber \\
{\frac{\beta_{10}}{\cH^{2}}} &=& 3\Omega_{m}f \\ \nonumber\\
{\frac{\beta_{11}}{\cH^{2}}} &=& {{{3\over2}\Omega_{m}\bigg(1+\frac{2f}{3\Omega_m}\bigg)}} + 3\Omega_mf - f^{2}\Bigg[-1+b_{e}-2\Q- \frac{2(1 {-} \Q)}{\chi\cH}-\frac{\cH'}{\cH^{2}}\Bigg] \\ \nonumber\\
{\frac{\beta_{12}}{\cH^{2}}} &=&{{-3\Omega_{m}\bigg(1+\frac{2f}{3\Omega_m}\bigg) }}
- f\Bigg[b_{{10}}(f-3+b_{e}) + \frac{b_{{10}}'}{\cH}\Bigg] \nonumber \\ &&{}
+ \frac{3}{2}\Omega_{m}\Bigg[b_{{10}}\bigg(2+b_{e}-4\Q-\frac{2(1-\Q)}{\chi\cH} -\frac{\cH'}{\cH^{2}}\bigg) + \frac{b_{{10}}'}{\cH} + 2\bigg(2-\frac{1}{\chi\cH}\bigg)\frac{\p b_{{10}}}{\p \ln{{L}}}\Bigg] 
  \\ \nonumber\\
\frac{\beta_{13}}{\cH^{2}} &=& \frac{9}{4}\Omega_{m}^{2} {+ \frac{3}{2}\Omega_{m}f\Bigg[1-2f+2b_{e}-{6}\Q-\frac{4(1-\Q)}{\chi\cH}-\frac{3\cH'}{\cH^{2}}\Bigg]} + f^{2}(3-b_{e}) \\ \nonumber\\
\frac{\beta_{14}}{\cH} &=& -\frac{3}{2}\Omega_{m}b_{{10}} \\\nonumber\\
\frac{\beta_{15}}{\cH} &=& 2f^{2}  \\ \nonumber\\
\frac{\beta_{16}}{\cH} &=& f\Bigg[b_{{10}}\bigg(f+b_{e}-2\Q-\frac{2(1-\Q)}{\chi\cH}-\frac{\cH'}{\cH^{2}}\bigg) + \frac{b_{{10}}'}{\cH} + 2\bigg(1-\frac{1}{\chi\cH}\bigg)\frac{\p b_{{10}}}{\p \ln {L}}\Bigg] \\\nonumber\\
\frac{\beta_{17}}{\cH} &=& -\frac{3}{2}\Omega_{m}f \\\nonumber\\
\frac{\beta_{18}}{\cH} &=& \frac{3}{2}\Omega_{m}f -f^{2}\Bigg[3-2b_{e}+{4}\Q+\frac{4(1-\Q)}{\chi\cH}+\frac{3\cH'}{\cH^{2}}\Bigg] \\\nonumber\\
\frac{\beta_{19}}{\cH} &=& f\Bigg[b_{e}-2Q-\frac{2(1-\Q)}{\chi\cH}-\frac{\cH'}{\cH^{2}}\Bigg] 
\end{eqnarray}


\begin{table}[!t] 
\centering 
\caption{ {Individual terms in the observed $\Delta_g^{(2)}(a,\bm{x})$ [see \eqref{eq:SecondorderNewtonian}, \eqref{odg2}] for $\fnl=0$ are shown in column 1. The related $\beta_I$ functions in \eqref{e2.23} are listed in column 2. The Fourier-space kernels ${\cal F}$  corresponding to column 1, given by 
$ \int \!{\ud \k'}\, {\cal F}(\k',\k-\k')\delta_{\rm T}(\k')\delta_{\rm T}(\k-\k')/(2\pi)^3$, are shown in column~3. Column 4 gives
the coefficients of the terms in $\Delta_g^{(2)}$ (column 1). The line-of-sight derivative is $\partial_\|=\bm{n}\!\cdot\!\bm{\nabla}$ and $\Phi=\Psi$.
The superscript (1) on first-order quantities has been omitted and N denotes Newtonian. This table updates the one in \cite{Jolicoeur:2017nyt}.}} \label{tabc1} 
\vspace*{0.5cm}
\tiny
\begin{tabular}{|c|c|c|c|} 
\hline 
&  &  & \\
TERM & $~~\beta~~$ & FOURIER KERNEL    & COEFFICIENT \\ 
&  &  & \\ \hline \hline
&  &  & \\
$\delta^{(2)}_{\mathrm{T,{N}}}$ & N & $F_{2}(\!\bm{k}_{1},\!\bm{k}_{2}\!)$ & $b_{10}$ \\ 
&  &  & \\
${\big(\delta_{\mathrm{T}}\big)^2}$ & N & 1 & $b_{20}$ \\ 
&  &  & \\
${s^2}$ & N & $S_2(\!\bm{k}_{1},\!\bm{k}_{2}\!)$ & $b_{s}$ \\ 
&  &  & \\
$\partial_{\parallel}^{2}V^{(2)}_{{\rm N}}$ & N & $f^{2}\cH \mu_{3}^{2}G_{2}(\!\bm{k}_{1},\!\bm{k}_{2}\!)$ & $-1/\cH$ \\ 
&  &  & \\
$\delta_{\mathrm{T}}\partial_{\parallel}^{2}V$ & N & $-f\cH \big(\mu_{1}^{2} + \mu_{2}^{2}\big)/2 $ & $-2b_{10}/\cH$ \\
&  &  & \\ 
$\partial_{\parallel}V \partial_{\parallel}\delta_{\mathrm{T}}$ & N\ & $-f\cH {{\mu_{1}\mu_{2}\big(k_{1}^{2} + k_{2}^{2}\big)}/{\big(2k_{1}k_{2}\big)}}$& $-2b_{10}/\cH$ \\ 
&  &  & \\
$\partial_{\parallel}V \partial_{\parallel}^{3}V $  & N & $f^{2} \cH^2{{\big(\mu_{1}\mu_{2}^{3}k_{2}^{2} + \mu_{2}\mu_{1}^{3}k_{1}^{2}\big)}/{\big(k_{1}k_{2}\big)}}$ & $ {2}/{\cH^{2}}$  \\
&  &  & \\ 
$\big[\partial_{\parallel}^{2}V\big]^{2}$  & N &$f^{2}\cH^2 \,\mu_{1}^{2}\mu_{2}^{2}$ & ${2}/{\cH^{2}}$ \\ 
&  &  & \\
\hline 
&  &  & \\
$\big(\Psi\big)^{2}$ & $\beta_{1}$ & ${9}\Omega_{m}^{2}\cH^{4}/{\big(4k_{1}^{2}k_{2}^{2}\big)}$ & $\mathcal{A}_{1}$ \\ 
&  &  & \\
${\Psi V}$ & $\beta_{1}$ & $-{3}\Omega_{m}\cH^{3}f/{\big(2k_{1}^{2}k_{2}^{2}\big)}$  & $\mathcal{A}_{2}$ \\ 
&  &  & \\
$VV' $ & $\beta_{1}$  & ${ f\cH^{3}\big(3\Omega_{m}-2f\big)/{\big(2k_{1}^{2}k_{2}^{2}\big)} }$ & $ (b_{e}-3)\cH$ \\ 
&  &  & \\
$\big(V\big)^{2}$ & $\beta_{1}$ & $f^{2} \cH^{2}/{\big(k_{1}^{2}k_{2}^{2}\big)}$ & $(b_{e}-3)^{2}\cH^2+{b_{e}'\cH} +(b_{e}-3){\cH'} $ \\ 
&  &  & \\
${V^{(2)}_{\rm GR}}$ & $\beta_1,\beta_2$ & $-3\Omega_{m}\cH^{3}\big[3 -2 E_{2}(\!\bm{k}_{1},\!\bm{k}_{2},\!\bm{k}_{3}\!)\big]/\big(4k_{1}^{2}k_{2}^{2}\big)$ & $(3-b_{e})\cH$ \\
&  &  & \\
${\Phi^{(2)}_{\rm GR}}$ & $\beta_1,\beta_2$ & $3\Omega_{m}\cH^{4}\big[{f-\mathcal{C}_1 + \mathcal{C}_1} E_{2}(\!\bm{k}_{1},\!\bm{k}_{2},\!\bm{k}_{3}\!)\big]/\big(2k_{1}^{2}k_{2}^{2}\big)$ & $1-b_e+2\Q+\mathcal{R}$ \\
&  &  & \\
${\Psi^{(2)}_{\rm GR}}$ & $\beta_1,\beta_2$ & $3\Omega_{m}\cH^{4}\big[{\mathcal{C}_1-3f+2f^2 +2 f} E_{2}(\!\bm{k}_{1},\!\bm{k}_{2},\!\bm{k}_{3}\!)\big]/\big(2k_{1}^{2}k_{2}^{2}\big)$ & $2\big(\Q - 1\big)$ \\
&  &  & \\
${\Psi^{(2)\prime}_{\rm GR}}$ & $\beta_1,\beta_2$ & $3\Omega_{m}\cH^{5}\big[{\mathcal{C}_2 +  \mathcal{C}_{3}} E_{2}(\!\bm{k}_{1},\!\bm{k}_{2},\!\bm{k}_{3}\!)\big]/\big(2k_{1}^{2}k_{2}^{2}\big)$ & ${1}/{\cH}$ \\
&  &  & \\
\hline 
&  &  & \\
$V\partial_{\parallel}V $ & $\beta_{3}$ & $ \mathrm{i}\,f^{2}\cH^{2}
{\big(\mu_{1}k_{1} + \mu_{2}k_{2}\big)}/{\big(2k_{1}^{2}k_{2}^{2}\big)}$ &  $\mathcal{A}_{3}$ \\ 
&  &  & \\
$\Psi\partial_{\parallel}V$ & $\beta_{3}$ & $-{\rm i}\,3f\Omega_{m}\cH^{3}\,{\big(\mu_{1}k_{1} + \mu_{2}k_{2}\big)}/{\big(4k_{1}^{2}k_{2}^{2}\big)}$ & $\mathcal{A}_{4}$ \\ 
&  &  & \\
$\Psi\partial_{\parallel}\Phi$ & $\beta_{3}$ & ${\rm i}\,9\Omega_{m}^{2}\cH^{4}{\big(\mu_{1}k_{1} + \mu_{2}k_{2}\big)}/{\big(8k_{1}^{2}k_{2}^{2}\big)}$ & ${2}(f-2+2\mathcal{Q})/{\cH}$  \\ 
&  &  & \\
\hline
&  &  & \\ 
${\partial_{\parallel}V^{(2)}_{\rm GR}}$ & $\beta_{4},\beta_{5}$ & $-\mathrm{i}\,3\Omega_{m}\cH^{3}\big[3 -2 E_{2}(\!\bm{k}_{1},\!\bm{k}_{2},\!\bm{k}_{3}\!)\big]\mu_3 k_3/\big(4k_{1}^{2}k_{2}^{2}\big)$ & $b_{e}-2Q-\mathcal{R}$ \\
&  &  & \\
\hline
&  &  & \\ 
${\Psi^{(2)}_{\rm N}=\Phi^{(2)}_{\rm N}}$ & $\beta_{6}$ & $-{3}\Omega_{m}{\cH^{2}}F_{2}(\!\bm{k}_{1},\!\bm{k}_{2}\!)/{\big(2k_{3}^{2}\big)}$ & $4\mathcal{Q}-1-b_{e}+\mathcal{R}$ \\
&  &  & \\ 
${\Psi_{\rm N}^{(2)\prime} = \Phi_{\rm N}^{(2)\prime}}$  & $\beta_{6}$ & $-{3}\Omega_{m}{\cH^{3}}(2f-1)F_{2}(\!\bm{k}_{1},\!\bm{k}_{2}\!)/{\big(2k_{3}^{2}\big)}$ & ${1}/{\cH}$ \\
&  &  & \\ 
\hline\
&  &  & \\ 
$V^{(2)}_{\rm N}$ & $\beta_{7}$ & $f{\cH}G_{2}(\!\bm{k}_{1},\!\bm{k}_{2}\!)/{k_{3}^{2}}$ & $(3-b_{e})\cH$ \\ 
&  &  & \\
\hline 
&  &  & \\
$\big(\partial_{\parallel}V\big)^{2} $ & $\beta_{8}$ & $-f^{2}\cH^{2}{\mu_{1}\mu_{2}}/{\big(k_{1}k_{2}\big)}$ & $\mathcal{A}_{5}$ \\
&  &  & \\ 
$\partial_{\parallel}V\partial_{\parallel}\Psi$ & $\beta_{8}$ & ${3}f\Omega_{m}\cH^{3}{\mu_{1}\mu_{2}/{\big(2k_{1}k_{2}\big)}}$ & ${2}(2-f-2\mathcal{Q})/{\cH}$ \\
&  &  & \\ 
\hline
&  &  & \\ 
${\partial_{\parallel}^{2}V^{(2)}_{\rm GR}}$ & $\beta_{9},\beta_{10}$ & $\mathrm{i}\,3\Omega_{m}\cH^{3}\big[3 -2 E_{2}(\!\bm{k}_{1},\!\bm{k}_{2},\!\bm{k}_{3}\!)\big]\mu_3^{2} k_3^{2}/\big(4k_{1}^{2}k_{2}^{2}\big)$ & $-1/\cH$ \\
&  &  & \\ 
\hline
&  &  & \\ 
$\partial_{i}V\,\partial^{i}V$ & $\beta_{11}$ & $-f^{2}\cH^{2}\,{\bm{k}_1\!\cdot\! \bm{k}_2}/{\big(k_{1}^{2}k_{2}^{2}\big)}$  & $ b_{e} -1- 2\mathcal{Q} - \mathcal{R}$ \\ 
&  &  & \\
$\partial_{i}V\partial^{i}\Psi$  &$\beta_{11}$ & $3f\Omega_{m}\cH^{3} \,{\bm{k}_{1}\!\cdot\! \bm{k}_{2}}/{\big(2k_{1}^{2}k_{2}^{2}\big)} $ & ${2}/{\cH}$  \\
&  &  & \\ 
\hline 
&  &  & \\
${\Psi\delta_{\mathrm{T}}}$ & $\beta_{12}$ &$-{3}\Omega_{m}\cH^{2}{\big(k_{1}^{2}+k_{2}^{2}\big)}/{\big(4k_{1}^{2}k_{2}^{2}\big)} $& $2b_{10}\big(4\mathcal{Q}+\mathcal{R} -2-b_{e}\big) -{\mathcal{S}}$  \\ 
&  &  & \\
${V \delta_{\mathrm{T}}}$& $\beta_{12}$ & $f\cH{\big(k_{1}^{2}+k_{2}^{2}\big)}/{\big(2k_{1}^{2}k_{2}^{2}\big)} $  & $b_{10}' + 2b_{10}\big(3-b_e-f\big)\cH$ \\ 
&  &  & \\
\hline 
\end{tabular}
\end{table}
\clearpage
\begin{table*}[!t] 
\centering 
\caption{Table \ref{tabc1}  continued.} \label{tabc2} 
\vspace*{0.5cm}
\tiny
\begin{tabular}{|c|c|c|c|} 
\hline 
&  &  & \\
TERM & $~~\beta~~$ & FOURIER KERNEL  & COEFFICIENT \\ 
&  &  & \\ \hline \hline
&  &  & \\
{$\delta^{(2)}_{{g \rm T,GR}}$}& ${\beta_{11},\beta_{12}}$ & {$\big(3\Omega_m+2f\big)\cH^2\big[\bm{k}_{1}\cdot\!\bm{k}_{2}-2\big(k_1^2+k_2^2\big)\big]/\big(2k_{1}k_{2}\big)$} & 1 \\
&  &  & \\ 
\hline
&  &  & \\
$\Psi \partial^{2}_{\parallel}V$ & $\beta_{13}$ & ${3}f\Omega_{m}\cH^{3}{\big(\mu_{1}^{2}k_{1}^{2} + \mu_{2}^{2}k_{2}^{2}\big)}/{\big(4k_{1}^{2}k_{2}^{2}\big)}$ &${2}\big[{1}-2f+2b_{e}-{6}\mathcal{Q}-2{\cal R}-\big({\cH'}/{\cH^{2}}\big)\big]/{\cH}$ \\ 
&  &  & \\
$\Psi\partial_{\parallel}^{2}\Psi $ & $\beta_{13}$ & $- {9}\Omega_{m}^2\cH^{4}{\big(\mu_{1}^{2}k_{1}^{2} + \mu_{2}^{2}k_{2}^{2}\big)}/{\big(4k_{1}^{2}k_{2}^{2}\big)}$ & $-{2}/{\cH^{2}}$  \\ 
&  &  & \\
$V\partial_{\parallel}^{2}V$ &$\beta_{13}$ & $-f^2\cH^{3}{\big(\mu_{1}^{2}k_{1}^{2} + \mu_{2}^{2}k_{2}^{2}\big)}/{\big(2k_{1}^{2}k_{2}^{2}\big)}$ & ${2}(b_{e}-3)/{\cH}$ \\ 
&  &  & \\
$\Psi\partial_{\parallel}\delta_{\mathrm{T}}$ &$\beta_{14}$ & $-{\rm i}\,3\Omega_{m}\cH^2 {\big(\mu_{1}k_{1}^{3} + \mu_{2}k_{2}^{3}\big)}/{\big(4k_{1}^{2}k_{2}^{2}\big)}$& ${2}b_{10}/{\cH}$ \\ 
&  &  & \\
\hline
&  &  & \\
$\partial_{i}V\partial_{\parallel}\partial^{i}V $ & $\beta_{15}$ &$ -\mathrm{i}\,f^2\cH^2 \bm{k}_{1}\!\cdot\! \bm{k}_{2}{\big(\mu_{1}k_{1} + \mu_{2}k_{2}\big)}/{\big(2k_{1}^{2}k_{2}^{2}\big)}$ & $-{4}/{\cH}$ \\ 
&  &  & \\
\hline 
&  &  & \\
${\delta_{\mathrm{T}}\partial_{\parallel}V}$ & $\beta_{16}$ & $\mathrm{i}\,f\cH{\big(\mu_{1}k_{2} + \mu_{2}k_{1}\big)}/{\big(2k_{1}k_{2}\big)}$ & $2b_{10}\big(f+b_e-2\Q-\mathcal{R}\big)+{\mathcal{S}}$  \\ 
&  &  & \\
\hline
&  &  & \\ 
$\Phi\partial_{\parallel}^{3}V$ & $\beta_{17}$ & ${\rm i}\,3f\Omega_{m}\cH^3 {\big(\mu_{1}^{3}k_{1}^{3} + \mu_{2}^{3}k_{2}^{3}\big)}/{\big(4k_{1}^{2}k_{2}^{2}\big)}$ & $-{2}/{\cH^{2}}$ \\ 
&  &  & \\
\hline 
&  &  & \\
$\partial_{\parallel}V\partial^{2}_{\parallel}V$ & $\beta_{18}$ & $-\mathrm{i}\,f^{2}\cH^2 {{\big(\mu_{1}\mu_{2}^{2}k_{2} +\mu_{2}\mu_{1}^{2}k_{1}\big) }/{\big(2k_{1}k_{2}\big)}}$ & ${2}\big[3-2b_{e}+{4}\mathcal{Q}+2{\cal R}+\big({\cH'}/{\cH^{2}}\big)\big]/{\cH}$ \\ 
&  &  & \\
$\partial_{\parallel}V\partial^{2}_{\parallel}\Psi$ & $\beta_{18}$ & ${\rm i}\,3f\Omega_{m}\cH^3{{\big(\mu_{1}\mu_{2}^{2}k_{2} +\mu_{2}\mu_{1}^{2}k_{1}\big) }/{\big(4k_{1}k_{2}\big)}}$ & ${2}/{\cH^{2}}$ \\ 
&  &  & \\
\hline 
&  &  & \\
$\partial_{\parallel}{V_{\rm N}^{(2)}}$& $\beta_{19}$ & $\mathrm{i}\,f\cH \,{\mu_{3}}G_{2}(\!\bm{k}_{1},\!\bm{k}_{2}\!)/{k_{3}}$ & $b_{e}-2Q-\mathcal{R}$ \\
&  &  & \\ 
\hline

\end{tabular}
\end{table*}
\noindent Here the ${\cal C}$ functions in the Fourier kernels are
\bea
{\mathcal{C}_{1}} &=& 2f - f^2 -3\Omega_m\;, \label{e7} \\
\mathcal{C}_{2} &=&2f-1+ \big(1-f\big)\bigg[6\Omega_m + f\big(1-2f\big) - 2f\frac{\cH'}{\cH^{2}}\bigg] \;, \label{e10} \\
\mathcal{C}_{3} &=&2 f\bigg(2f-1+\frac{\cH'}{\cH^{2}}\bigg) +2 \frac{f'}{\cH}\;, \label{e12} 
\eea
the 
${\cal A}$ functions in the coefficients are 
\begin{eqnarray}
\mathcal{A}_{1} &=& {-3} +2f\bigg({2}-2b_{e}+{4}\mathcal{Q}+\frac{4(1-\mathcal{Q})}{\chi\cH} +\frac{2\cH'}{\cH^{2}}\bigg) -\frac{2f'}{\cH} +b_{e}^{2}+ 6b_{e}-8b_{e}\mathcal{Q}+4\mathcal{Q} \nonumber \\
&& +16\mathcal{Q}^{2} -16\frac{\partial \mathcal{Q}}{\partial \ln{L}} -8\frac{\mathcal{Q}'}{\cH} + \frac{b_{e}'}{\cH}+\frac{2}{\chi^{2}\cH^{2}}\bigg(1-\mathcal{Q}+2\mathcal{Q}^{2}-2\frac{\partial \mathcal{Q}}{\partial \ln{{L}}}\bigg)  \nonumber\\ 
&& - \frac{2}{\chi\cH}\bigg[4+2b_{e}-2b_{e}\mathcal{Q}-4\mathcal{Q}+8\mathcal{Q}^{2}-\frac{3\cH'}{\cH^{2}}(1-\mathcal{Q})- 8\frac{\partial \mathcal{Q}}{\partial \ln{{L}}} - 2\frac{\mathcal{Q}'}{\cH}\bigg] \nonumber \\
&& + \frac{\cH'}{\cH^{2}}\bigg(-8-2b_{e}+{8}\mathcal{Q}+\frac{3\cH'}{\cH^{2}}\bigg) - \frac{\cH''}{\cH^{3}}\;, \label{e2} 
\eea
\bea 
\mathcal{A}_{2} &=& 2\cH\bigg[-\frac{15}{2}+f(3-b_{e})-\frac{3}{2}b_{e}-2b_{e}\frac{(1-\mathcal{Q})}{\chi\cH}+\frac{b_{e}'}{\cH}+b_{e}^{2}-4b_{e}\mathcal{Q}+12\mathcal{Q}+\frac{6(1-\mathcal{Q})}{\chi\cH} \nonumber \\
&& \qquad -2\bigg(2-\frac{1}{\chi\cH}\bigg)\frac{\mathcal{Q}'}{\cH}\bigg]\;, \label{e4} \\
\mathcal{A}_{3} &=& 2\cH\bigg[-3+4b_{e}+\frac{2b_{e}(1-\mathcal{Q})}{\chi\cH}-b_{e}^{2}+2b_{e}\mathcal{Q} -6\mathcal{Q}-\frac{b_{e}'}{\cH}-\frac{6(1-\mathcal{Q})}{\chi\cH} \nonumber \\
&& \qquad +2\bigg(1-\frac{1}{\chi\cH}\bigg)\frac{\mathcal{Q}'}{\cH}\bigg]\;, \label{e3} 
\\ 
\mathcal{A}_{4} &=& 4 +2f\bigg[-3+f+2b_{e}-3\mathcal{Q}-\frac{4(1-\mathcal{Q})}{\chi\cH}-\frac{2\cH'}{\cH^{2}}\bigg] +\frac{2f'}{\cH}-6b_{e}-2b_{e}^{2}+12b_{e}\mathcal{Q}-{8}\mathcal{Q} \nonumber\\
&& -16\mathcal{Q}^{2}+16\frac{\partial \mathcal{Q}}{\partial \ln{{L}}} +12\frac{\mathcal{Q}'}{\cH} -2\frac{b_{e}'}{\cH} -\frac{4}{\chi^{2}\cH^{2}}\bigg(1-\mathcal{Q}+2\mathcal{Q}^{2}-2\frac{\partial \mathcal{Q}}{\partial \ln{{L}}}\bigg)  \nonumber \\
&& - \frac{4}{\chi \cH}\bigg(-1 -2b_{e}+2b_{e}\mathcal{Q}+\mathcal{Q}-6\mathcal{Q}^{2}+\frac{3\cH'}{\cH^{2}}(1-\mathcal{Q}) +6\frac{\partial \mathcal{Q}}{\partial \ln{{L}}} + 2\frac{\mathcal{Q}'}{\cH}\bigg)  \nonumber \\
&& + \frac{2\cH'}{\cH^{2}}\bigg(3+2b_{e}-6\mathcal{Q}-\frac{3\cH'}{\cH^{2}}\bigg) + \frac{2\cH''}{\cH^{3}}\;, \label{e5}\\
\mathcal{A}_{5} &=&  -4-b_{e}+b_{e}^{2}-4b_{e}\mathcal{Q}+ {6}\mathcal{Q}+4\mathcal{Q}^{2}-4\frac{\partial \mathcal{Q}}{\partial \ln{{L}}} - 4\frac{\mathcal{Q}'}{\cH} + \frac{b_{e}'}{\cH} \nonumber \\
&& + \frac{2}{\chi^{2}\cH^{2}}\bigg(1-\mathcal{Q}+2\mathcal{Q}^{2}-2\frac{\partial \mathcal{Q}}{\partial \ln{{L}}}\bigg) \nonumber \\
&& + \frac{2}{\chi\cH}\bigg[3-2b_{e}+2b_{e}\mathcal{Q}-3\mathcal{Q}-4\mathcal{Q}^{2}+\frac{3\cH'}{\cH^{2}}(1-\mathcal{Q}) + 4\frac{\partial \mathcal{Q}}{\partial \ln{{L}}} + 2\frac{\mathcal{Q}'}{\cH}\bigg]  \nonumber \\
&& + \frac{\cH'}{\cH^{2}}\bigg(3-2b_{e}+{4}\mathcal{Q}+\frac{3\cH'}{\cH^{2}}\bigg) - \frac{\cH''}{\cH^{3}}\;, \label{e6} 
\end{eqnarray}
and the functions ${\cal R}, {\cal S}$ in the coefficients are
\begin{eqnarray}
\mathcal{R} &=& \frac{2(1-\mathcal{Q})}{\chi\cH}+\frac{\cH'}{\cH^{2}}\;, \label{e1}\\
\mathcal{S} &=& 4\bigg(2-\frac{1}{\chi\cH}\bigg)\frac{\partial b_{10}}{\partial \ln{{L}}}\;. \label{e1'}
\end{eqnarray}

The magnification bias is defined by \cite{Alonso:2015uua,DiDio:2015bua,Maartens:2019yhx}:
\be
{\cal Q}=- {\partial \ln \bar{n}_g \over \partial \ln L}\Bigg|_{\rm c}\,,
\ee
where $L$ is the background luminosity and the derivative is evaluated at the flux cut. Similarly,  $\partial b_{10} / \partial \ln L$ is understood to be evaluated at the flux cut.
We use a short-hand notation for the second luminosity derivative of $\bar{n}_g$:
\be
{\partial {\cal Q} \over \partial \ln L}\equiv - {\partial^2 \ln \bar{n}_g \over \partial (\ln L)^2}\Bigg|_{\rm c}\,.
\ee

\newpage
\section{{$\Upsilon_I$ functions in \eqref{e2.35}}}\label{app2}

\begin{eqnarray}
{1\over\fnl}\,\frac{\Upsilon_{1}}{{\cH^2}} &=& 2(3-b_e)f+ 3\Omega_m\bigg[1+b_e-4{\cal Q}-\frac{2(1-\mathcal{Q})}{\chi\cH}-\frac{\cH'}{\cH^{2}}\bigg]
\notag \\ &&{} {+{6\Omega_m \over \big(3\Omega_m+2f\big)}\,\bigg[{f'\over \cH}+\bigg(1+2{\cH' \over \cH^2} \bigg)f \bigg] }
\label{e2.26}
\\ &&\notag  \\
{1\over\fnl}\, \frac{\Upsilon_{2}}{{\cH}}& =& 
{2f \bigg[b_{e}-2\Q-\frac{2(1-\Q)}{\chi \cH} - \frac{\cH'}{\cH^{2}}\bigg]} \label{e2.27}
\\ && \notag \\
\frac{1}{b_{01}}\frac{{\Upsilon_{{3}}}}{\cH^{2}} &=& \frac{3}{2}\Omega_m\bigg[2+b_e-4\Q+\frac{2(1-\Q)}{\chi \cH} + \frac{\cH'}{\cH^{2}} + 2\bigg(2-\frac{1}{\chi \cH}\bigg)\frac{\partial \ln{b_{01}}}{\partial \ln{{L}}}\bigg] \nonumber \\
&& + f\bigg[3-f-b_e + \frac{1}{2}\frac{\partial \ln{b_{01}}}{\partial \ln{a}}\bigg] \label{e2.32} 
\\&& \notag  \\
{1\over b_{01}}\,\frac{{\Upsilon_{{4}}}}{\cH} &=& -\frac{3}{2}\Omega_{m}   \label{e2.33} 
\\&& \notag  \\
\frac{1}{b_{01}}\frac{{\Upsilon_{{5}}}}{\cH} &=& f\bigg[f+b_e-2\Q-\frac{2(1-\Q)}{\chi \cH}-\frac{\cH'}{\cH^{2}} + 2\bigg(2-\frac{1}{\chi \cH}\bigg)\frac{\partial \ln{b_{01}}}{\partial \ln{{L}}}\bigg]\label{e2.34}
\\&& \notag
\end{eqnarray}
Note that $\Upsilon_2= 2 \fnl\,\gamma_1$.


\begin{table}[!h] 
\centering 
\caption{{The $\fnl \neq 0$ terms from relativistic projection effects [see \eqref{e2.35}].}} \label{tabc3} 
\vspace*{0.5cm}
\tiny
\begin{tabular}{|c|c|c|c|} 
\hline 
&  &  & \\
TERM & $~~\Upsilon~~$ & FOURIER KERNEL  & COEFFICIENT \\ 
&  &  & \\ \hline \hline
&  &  & \\ 
$V^{(2)}_{\rm {nG}}$ & $\Upsilon_{1}$ & ${2\fnl\, \cH f {\cal M}_3/\big({\cal M}_1 {\cal M}_2 k_3^2\big)}$ & $(3-b_{e})\cH$ \\ 
&  &  & \\
$\Psi^{(2)}_{\rm {nG}}=\Phi^{(2)}_{\rm {nG}}$ & $\Upsilon_{1}$ & ${-3\fnl\Omega_m \cH^{2}{\cal M}_3/\big({\cal M}_1 {\cal M}_2 k_3^2\big)}$ & ${4{\cal Q}-1-b_e+{\cal R} }$ \\ 
&  &  & \\
${\Psi^{(2)\prime}_{\rm nG}}$ & $\Upsilon_{1}$ & ${6\fnl\big[f' +\big(\cH+2\cH'/\cH \big)f \big]\Omega_m \cH^{2}{\cal M}_3/\big[\big( 3\Omega_m+2f\big) \big({\cal M}_1 {\cal M}_2 k_3^2\big)\big]}$ & $1/\cH$ \\ 
&  &  & \\
\hline 
&  &  & \\
$ \partial_{\parallel}V^{(2)}_{\rm {nG}}$ & $\Upsilon_2$ & $\mathrm{i}\;{2\fnl\, \cH f \,\mu_3{\cal M}_3/\big({\cal M}_1 {\cal M}_2 k_3\big)}$ & $b_{e}-2Q-\mathcal{R}$ \\ 
&  &  & \\
\hline 
&  &  & \\
$\Psi {\varphi_{\rm p}}$ & ${\Upsilon_{{3}}}$ & $-3\Omega_m \cH^{2}\big[\big(k_{1}^{2}/\mathcal{M}_{1}\big)+\big(k_{2}^{2}/\mathcal{M}_{2}\big)\big]/{\big(4k_{1}^{2}k_{2}^{2}\big)}$ & 
${ b_{01}\big[8\mathcal{Q}+2\mathcal{R} -2b_{e}-4 -\mathcal{S}/\big(b_{10}-1\big)\big]}$ \\ 
&  &  & \\
$V{\varphi_{\rm p}}$ & ${\Upsilon_{{3}}}$ & $f\cH\big[\big(k_{1}^{2}/\mathcal{M}_{1}\big)+\big(k_{2}^{2}/\mathcal{M}_{2}\big)\big]/{\big(2k_{1}^{2}k_{2}^{2}\big)}$ & ${b_{01}\big[ 2\big(3-b_e-f\big)\cH+b_{10}'/(b_{10}-1) \big]}$ \\ 
&  &  & \\
\hline 
&  &  & \\
$\Psi \partial_{\parallel}{\varphi_{\rm p}}$& ${\Upsilon_{{4}}}$ & $-\mathrm{i}\,3\Omega_m \cH^{2}\big[\big(\mu_1 k_{1}^{3}/\mathcal{M}_{1}\big)+\big(\mu_2 k_{2}^{3}/\mathcal{M}_{2}\big)\big]/{\big(4k_{1}^{2}k_{2}^{2}\big)}$ & ${2}b_{01}/{\cH}$ \\
&  &  & \\
\hline 
&  &  & \\
${\varphi_{\rm p}}\partial_{\parallel} V$ & ${\Upsilon_{{5}}}$ & $\mathrm{i}\,f\cH\big[\big(\mu_1 k_{2}/\mathcal{M}_{2}\big)+\big(\mu_2 k_{1}/\mathcal{M}_{1}\big)\big]/{\big(2k_{1}k_{2}\big)}$ & 
${ b_{01}\big[2f+2b_e-4\mathcal{Q}-2\mathcal{R} +\mathcal{S}/\big(b_{10}-1\big)\big]}$ \\
&  &  & \\
\hline
\end{tabular}
\end{table}

\clearpage
\bibliographystyle{JHEP}
\bibliography{reference_library}

\end{document}